\documentclass[prc,twocolumn,nofootinbib,superscriptaddress,showpacs,aps]{revtex4-1}

\usepackage{graphicx,amsmath,amssymb,bm}

\begin{document}

\title{Uncertainties in constraining low-energy constants from $^3$H $\beta$ decay}

\author{P.\ Klos}
\email[E-mail:~]{pklos@theorie.ikp.physik.tu-darmstadt.de}
\affiliation{Institut f\"ur Kernphysik, 
Technische Universit\"at Darmstadt, 
64289 Darmstadt, Germany}
\affiliation{ExtreMe Matter Institute EMMI, 
GSI Helmholtzzentrum f\"ur Schwerionenforschung GmbH, 
64291 Darmstadt, Germany}
\author{A.\ Carbone}
\email[E-mail:~]{arianna@theorie.ikp.physik.tu-darmstadt.de}
\affiliation{Institut f\"ur Kernphysik, 
Technische Universit\"at Darmstadt, 
64289 Darmstadt, Germany}
\affiliation{ExtreMe Matter Institute EMMI, 
GSI Helmholtzzentrum f\"ur Schwerionenforschung GmbH, 
64291 Darmstadt, Germany}
\author{K.\ Hebeler}
\email[E-mail:~]{kai.hebeler@physik.tu-darmstadt.de}
\affiliation{Institut f\"ur Kernphysik, 
Technische Universit\"at Darmstadt, 
64289 Darmstadt, Germany}
\affiliation{ExtreMe Matter Institute EMMI, 
GSI Helmholtzzentrum f\"ur Schwerionenforschung GmbH, 
64291 Darmstadt, Germany}
\author{J.\ Men\'{e}ndez}
\email[E-mail:~]{menendez@nt.phys.s.u-tokyo.ac.jp}
\affiliation{Department of Physics, University of Tokyo, Hongo, 
Tokyo 113-0033, Japan}
\author{A.\ Schwenk}
\email[E-mail:~]{schwenk@physik.tu-darmstadt.de}
\affiliation{Institut f\"ur Kernphysik, 
Technische Universit\"at Darmstadt, 
64289 Darmstadt, Germany}
\affiliation{ExtreMe Matter Institute EMMI, 
GSI Helmholtzzentrum f\"ur Schwerionenforschung GmbH, 
64291 Darmstadt, Germany}
\affiliation{Max-Planck-Institut f\"ur Kernphysik, 
Saupfercheckweg 1, 
69117 Heidelberg, 
Germany}

\begin{abstract}
We discuss the uncertainties in constraining low-energy constants of chiral
effective field theory from $^3$H $\beta$ decay. The half-life is very
precisely known, so that the Gamow-Teller matrix element has been used to fit
the coupling $c_D$ of the axial-vector current to a short-range two-nucleon
pair. Because the same coupling also describes the leading one-pion-exchange
three-nucleon force, this in principle provides a very constraining fit,
uncorrelated with the $^3$H binding energy fit used to constrain another
low-energy coupling in three-nucleon forces. However, so far such $^3$H
half-life fits have only been performed at a fixed cutoff value. We show that
the cutoff dependence due to the regulators in the axial-vector two-body
current can significantly affect the Gamow-Teller matrix elements
and consequently also the extracted values for the $c_D$ coupling constant.
The degree of the cutoff dependence is correlated with the softness of the
employed NN interaction. As a result, present three-nucleon forces based
on a fit to $^3$H $\beta$ decay underestimate the uncertainty in $c_D$. We
explore a range of $c_D$ values that is compatible within cutoff variation
with the experimental $^3$H half-life and estimate the resulting uncertainties
for many-body systems by performing calculations of symmetric nuclear matter.
\end{abstract}


\maketitle

{\it Introduction.} The development of new and improved nuclear forces within
chiral effective field theory (EFT) is currently a very active field of
research~\cite{Geze13QMCchi,Ekst15sat,Epel15NNn4lo,Carl15sim}. In contrast to
phenomenological approaches, chiral EFT provides a framework that allows to
systematically derive improvable expansions for nucleon-nucleon (NN) and
many-body forces as well as electroweak current operators at low
energies~\cite{Epel09RMP,Mach11PR,Hamm13RMP,Bacca14JPG}. Within Weinberg's
power counting scheme~\cite{Wein91chNp} the contributions to NN forces have
been worked out up to fifth order in the chiral
expansion~\cite{Epel15NNn4lo,Ente14NNn4lo}, whereas three-nucleon (3N) and
four-nucleon forces have been developed up to fourth
order~\cite{Ishi073N2p,Bern083Nlong,Bern113Nshort,Epel064N}. Similarly nuclear
currents have been derived up to fourth order for
axial-vector~\cite{Kreb16axial,Baro15axial} and vector currents~
\cite{Koll09curr,Koll11em2b,Past09EMcurr,Past11EMcharge,Piar13EMcurr}.

The contributions to nuclear forces and currents generally depend on
low-energy couplings (LECs) that capture the short-distance physics that is
not resolved explicitly within the EFT. Therefore, to determine the LECs,
fits to data are needed. Up to second (next-to-leading) order there are no
contributions from many-body interactions or currents, and the LECs in the NN
forces are usually obtained by fits to pion-nucleon and nucleon-nucleon
scattering data. At third (next-to-next-to leading) order, N$^2$LO, two
additional LECs, $c_D$ and $c_E$, enter in 3N forces and two-body (2b)
currents. While $c_D$ enters in both the NN-contact-one-pion exchange 3N force
and the coupling of a NN pair to an axial-vector probe, $c_E$ only
parameterizes the 3N short-range contact interaction.

Different strategies have been employed to determine the values of the 3N LECs
up to N$^2$LO. If only the new couplings appearing at this order are to be
fixed, two observables are needed. One popular choice, first introduced in
Ref.~\cite{Gazi08lec}, is to determine $c_D$ and $c_E$ by fits to the binding
energy of $^3$H (triton) or $^3$He, and the $^3$H $\beta$-decay half-life.
This procedure, also used in other studies~\cite{Marc12mucap,Ekst14GT2bc}, has
recently been extended by including selected terms up to fifth order in the
nuclear currents~\cite{Baro16beta}. On the other hand, in
Ref.~\cite{Hebe11fits} $c_D$ and $c_E$ were fixed by fits to the $^3$H binding
energy and the $^4$He charge radius. Moreover, different pairs of data have
been chosen to determine $c_D$ and $c_E$, such as the $^3$H binding energy and
the neutron-deuteron scattering length~\cite{Epel02fewbody}, or the
$^4$He binding energy and the $P$-wave spin-orbit splitting in neutron-$^4$He
scattering~\cite{Lynn16QMC3N}. An alternative strategy was employed in
Ref.~\cite{Carl15sim}, following similar ideas as in Ref.~\cite{Ekst15sat}.
Instead of fitting the NN and 3N interactions separately, all LECs up to a
given order in the chiral expansion were fit simultaneously based on a
$\chi^2$ minimization in a large parameter space. The correlations between
different LECs were also studied systematically.

Generally, reliable fits require observables that are not strongly correlated
under variations of the LECs. In particular, for the $c_D$ and $c_E$
determination, Ref.~\cite{Gazi08lec} showed that this condition is fulfilled
for fits to the $^3$H half-life and binding energy, as proposed
earlier~\cite{Gard06weak3nf}. With the LECs fixed, heavier systems can be
studied. The interactions based on Ref.~\cite{Gazi08lec} provide good nuclear
structure properties, including binding energies, up to oxygen
isotopes~\cite{Barr13PPNP,Hebe15ARNPS,Navr16PS}, but tend to overbind heavier
nuclei \cite{Herg13IMSRG,Bind14CCheavy,Herg14MR,
Hebe15ARNPS}. The significance of the overbinding is unclear, however, because the
uncertainties associated to the interactions have not been systematically
explored. Other interactions fitted to different data lead to promising
results for heavier nuclei, neutron and nuclear matter~\cite{Hu16nm5,Hebe11fits,
Hebe15ARNPS,Ekst15sat,Hage16NatPhys,Simo16unc,Simo17SatFinNuc,Dris16asym,Ruiz16Calcium,Hage16Ni78,
Lynn16QMC3N,Birk16edp}.

In this work, we investigate the theoretical uncertainties of LEC
determinations involving the $^3$H half-life and study to what extent the
LECs can be constrained based on these fits. Presently, a major source of
uncertainty that has not yet been properly taken into account is the
regularization scheme and scale dependence for both nuclear interactions and
currents. So far, two different regulators in momentum space have been used:
first, local regulators that affect all matrix elements in the momentum
transfer; and second, nonlocal regulators that act on the relative momenta in
the initial and final states. The consistent and efficient choice of a
regulator scheme is subject of an active ongoing debate as various different
schemes and scales are currently used for nuclear interactions~\cite{Ente03EMN3LO,
Navr07local3N,Geze13QMCchi,Carl15sim,Epel15NNn4lo}. Given that nuclear
structure observables are already sensitive to a specific choice (see,
e.g., Refs.~\cite{Tews16QMCPNM,Dyhd16Regs}) the consistent treatment of
regulators in interactions and currents represents an additional
challenge. First studies concerned with the uncertainty due to the choice of
regulators were performed only for a
relatively small range of cutoff values~\cite{Cora14nmat, Samm15numat}. In
this paper we show that the regulator choice can affect significantly the LEC values
extracted from $^3$H $\beta$ decay. This finding raises the fundamental
question regarding the consistency of the
regularization scheme in nuclear interactions and currents.

{\it Triton $\beta$ decay and nuclear currents.} Formally, the half-life $t$
for the $\beta$ decay of $^3$H can be expressed in the
form~\cite{Schi98weakcap,Rama78tables}
\begin{equation} (1+\delta_R) t = \frac{K/G_V^2}{f_V \, \langle \text{F}
\rangle^2
+ f_A \, g_A^2 \, \langle \text{GT} \rangle^2} \,,
\label{eq:half-life}
\end{equation} where $\delta_R$ includes radiative corrections that originate
from virtual photon exchange between the charged particles, $f_V$ and $f_A$
are Fermi functions, which account for the deformation of the electron wave
function due to electromagnetic interactions with the nucleus, and $G_V=1$ and
$g_A=1.27$ denote the vector and axial-vector couplings. The kinematics of the
process leads to an additional constant $K= 2\pi^3\ln 2 /m_e^5$, where $m_e$
is the electron mass. The half-life depends on the nuclear matrix elements of
the vector and axial-vector currents denoted as Fermi $\langle
\text{F}\rangle$ and Gamow-Teller $\langle \text{GT}\rangle$ matrix elements,
respectively. The Fermi reduced matrix element is given by $\left< \text{F}
\right> = \langle ^3\text{He} \lVert \sum^3_{i=1} \tau_i^+ \rVert
^3\text{H}\rangle\,$, where $\tau^+=\frac{1}{2}(\tau^x+i\tau^y)$ is the
isospin-raising operator. The Gamow-Teller reduced matrix element contains
axial-vector one-body (1b) and 2b current contributions:
\begin{equation}
\langle \text{GT} \rangle = \frac{1}{g_A} \langle ^3\text{He} \lVert
\sum^3_{i=1}{\bf J}^+_{i,{\rm 1b}}+\sum_{i<j}{\bf J}_{ij,\text{2b}}^+ \rVert
^3\text{H}\rangle \,.
\label{eq:gt_me}
\end{equation}

The axial-vector current was derived in chiral EFT to third
order~\cite{Park02eftsolar,Ando02mud,Hofe15powerdm}, also denoted as $Q^3$, where $Q\sim
m_{\pi}$ is the typical momentum scale, of the order of the pion mass, and the
expansion refers to powers in $Q/\Lambda_b$, with $\Lambda_b\sim 500$~MeV the
breakdown scale of the EFT.  More recent derivations have been performed to
order $Q^4$~\cite{Kreb16axial,Baro15axial}. Since the Q-value of the
$^3$H $\beta$ decay is about 100~keV, to very good approximation we evaluate
all currents at vanishing momentum transfer. Therefore to order
$Q^0$ and $Q^2$, only the momentum-independent 1b current contributes:
\begin{equation} {\bf J}^+_{i,{\rm 1b}} = g_A \tau_i^+ {\bm \sigma}_i \,.
\label{eq:GT_onebody}
\end{equation}

At order $Q^3$, 2b currents enter. In the limit of vanishing momentum
transfer, they are given by ~\cite{Park02eftsolar,Hofe15powerdm}
\begin{align} &{\bf J}_{12,\text{2b}}^+=-\frac{g_A}{2F_\pi^2}\frac{1}{{\bf
k}^2+m_\pi^2}\Bigl[4c_3{\bf k}\,{\bf k}\cdot(\tau_1^+ {\bm \sigma}_1+\tau_2^+
{\bm \sigma}_2)\nonumber\\ &\quad +\Bigl(c_4+\frac{1}{4m_N}\Bigr)({\bm
\tau}_1\times{\bm \tau}_2)^+\,{\bf k}\times\left[({\bm \sigma}_1\times{\bm
\sigma}_2)\times{\bf k}\right]\nonumber\\ &\quad -\frac{i}{8m_N}({\bm
\tau}_1\times{\bm \tau}_2)^+({\bf p}_1+{\bf p}'_1-{\bf p}_2-{\bf
p}'_2)({\bm\sigma}_1-{\bm \sigma}_2)\cdot{\bf k}\Bigr]\nonumber\\
&\quad-2id_1(\tau^+_1{\bm \sigma}_1+\tau^+_2{\bm
\sigma}_2)-id_2({\bm\tau}_1\times{\bm\tau}_2)^+({\bm\sigma}_1
\times{\bm\sigma}_2)\,,
\label{Eq_axial_curr_q0}
\end{align} where $(\tau_1 \times \tau_2)^+=(\tau_1 \times \tau_2)^x+i(\tau_1
\times
\tau_2)^y$, ${\bf k}=({\bf k}_2-{\bf k}_1)/2$, ${\bf k}_i={\bf p}'_i-{\bf
p}_i$, with initial and final nucleon momenta ${\bf p}_i$ and ${\bf p}_i'$,
and the pion decay constant $F_\pi=92.4$~MeV. We use $m_\pi=138$~MeV and for
the nucleon mass $m_N=938.9$~MeV. The $c_i$ are pion-nucleon LECs, which we
take consistently with the corresponding nuclear interaction.

The relativistic corrections of the leading 2b currents are suppressed by a
factor of $1/m_N$. In our counting this factor leads to an increase of the
chiral order by two units since $Q/m_N\sim(Q/\Lambda_b)^2$.  Therefore,
relativistic corrections to 2b currents are of order $Q^4$. Nevertheless we
include the $1/(4m_N)$ correction term of $c_4$ in
Eq.~\eqref{Eq_axial_curr_q0} in order to be consistent with
Ref.~\cite{Gazi08lec}, but its effect is minor. Similarly, the contribution of
the term proportional to $i/(8m_N)$ is only about $0.001\%$ of the total
Gamow-Teller matrix element and can thus be neglected.

Antisymmetrization of the short-range part of the 2b currents allows to
express $d_1$ and $d_2$ in terms of one linear combination $c_D=\Lambda_\chi
F_\pi^2 (d_1+2d_2)$ with
$\Lambda_\chi=700$~MeV~\cite{Park02eftsolar}.\footnote{Note that this is no
longer strictly the case for local regulators.} The coupling $c_D$ describes
the strength of a pion or an axial-vector current interacting with a
short-range NN pair. This relation is a generalization of the
Goldberger-Treiman relation to the 2b level~\cite{Gard06weak3nf}. The total
strength $d_R$ of the short-range part includes contributions from $c_3$,
$c_4$ and the minor relativistic correction:
\begin{equation} d_R=\frac{1}{\Lambda_\chi
g_A}c_D+\frac{1}{3}(c_3+2c_4)+\frac{1}{6m_N}\,.
\label{eq:dR_def}
\end{equation}

\begin{figure}[t]
\begin{center}
\includegraphics[width=0.45\textwidth,clip=]{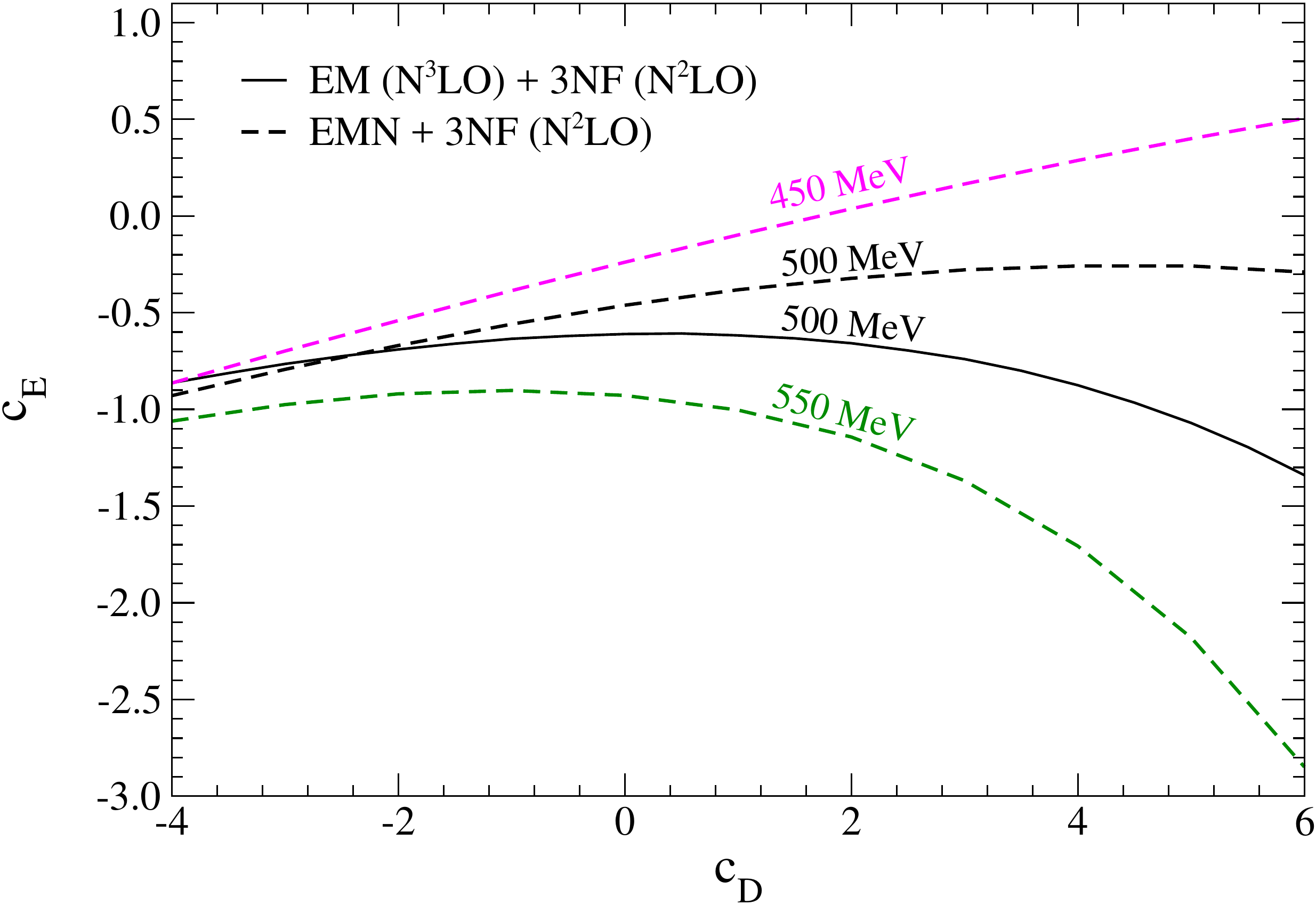}
\end{center}
\caption{Values for the low-energy couplings $c_D$ and $c_E$ that reproduce
the $^3$H binding energy using the EM 500 MeV N$^3$LO potential of
Ref.~\cite{Ente03EMN3LO} plus 3N interactions at order N$^2$LO
(solid), and using the EMN potentials of Ref.~\cite{Ente17EMn4lo} plus consistent 3N interactions 
at order N$^2$LO (dashed) for different cutoff values.}
\label{fig:3Hfits}
\end{figure}

{\it Nuclear states and regulators.} We evaluate the Fermi and Gamow-Teller
matrix elements in a momentum-space partial-wave representation. For this we
calculate $^3$H and $^3$He by solving the Faddeev equations in a partial-wave
momentum basis using different chiral NN interactions at both N$^2$LO and N$^3$LO plus 3N
interactions at N$^2$LO. Specifically, we use a $Jj$-coupled three-body basis
of the form
\begin{equation} | p q \alpha  \rangle = \left| p q; \left[ (L S) J (l s) j
\right] \mathcal{J} \mathcal{J}_z (T t)
\mathcal{T} \mathcal{T}_z \right> \, ,
\label{eq:JJbas}
\end{equation} where $L$, $S$, $J$ and $T$ denote the relative orbital angular
momentum, spin, total angular momentum and isospin of particles 1 and 2 with
relative momentum $p$. The single-particle quantum numbers $l$, $s=1/2$, $j$
and $t=1/2$ label the orbital angular momentum, spin, total angular momentum
and isospin of particle $3$ with momentum $q$ relative to the center-of-mass
of particles 1 and 2.  In this paper all calculations for $^3$H and $^3$He are
performed using the averaged neutron-proton mass $m=1/2(m_n + m_p)$ for all
three particles. This approximation is known to provide binding energy results
within 10 keV of the full results~\cite{Nogg03triton}. In this approximation
the three-body total angular momentum and isospin of the ground states take
the values $\mathcal{J} = \mathcal{T} = 1/2$, with magnetic projections
$\mathcal{J}_z$ and $\mathcal{T}_z$, and we include all partial waves up to
two-body $J_{\text{max}} = 6$.

For the calculation of the nuclear states we fix the value of $c_E$ for a
given value of $c_D$ by fitting the binding energy of $^3$H to the
experimental value $E_{^3\text{H}} = -8.482$~MeV. The resulting LEC values are
shown in Fig.~\ref{fig:3Hfits} for the calculation based on the NN interaction
at N$^3$LO of Ref.~\cite{Ente03EMN3LO} (EM 500 MeV) combined with 3N forces at
N$^2$LO (solid line). Here we used a non-local three-body regulator of the
form $f{}_{\text{3N}} (p,q,n_{\text{exp}}) = \exp[-((p^2 + 3/4
q^2)/\Lambda_{\rm 3N}^2)^{n_{\text{exp}}}]$ with $\Lambda_{\text{3N}} =
500$~MeV and $n_{\text{exp}} = 3$. Furthermore, we performed fits based on the
potentials of Ref.~\cite{Ente17EMn4lo} (EMN) at order N$^2$LO 
including the consistent 3N (with same $c_i$) force contributions at this order for cutoff
values o{}f 450~MeV, 500~MeV, and 550~MeV (dashed lines) and
$n_{\text{exp}}=4$. 
Figure~\ref{fig:3Hfits} shows that
the $c_E$ values are of natural size for the entire range of $c_D$ values and
all employed NN interactions.

The choice of regulator should be considered carefully when studying
transition operators such as $\beta$ decay. Generally, any nuclear interaction
contains intrinsic resolution scales $\Lambda$ that separate the low-energy
degrees of freedom that are treated explicitly from the high-energy degrees of
freedom that are contained implicitly in the coupling constants. As a result,
the $^3$H and $^3$He wave functions contain these resolution scales and are
hence suppressed at momenta that lie well beyond them. For the fit results
shown in Fig.~\ref{fig:3Hfits} the cutoff scales $\Lambda_\text{3N}$ were chosen to
be the same as in the corresponding NN interactions. 
When evaluating expectation values of operators with respect to these wave functions the
fundamental question arises on how the high-momentum components of the
operators need to be regularized. In Ref.~\cite{Gazi08lec} the 2b current
operators were regularized by local regulators of the form
\begin{equation} f_\Lambda^{\text{loc}}({\bf p},{\bf p}')=\exp \left[ -({\bf
p}-{\bf p}')^4/\Lambda^4 \right],
\label{eq:local_reg}
\end{equation}
where $\mathbf{p}$ and $\mathbf{p}'$ denote the relative momenta in the initial and final state.
In contrast the NN interactions derived in
Refs.~\cite{Ente03EMN3LO,Epel05EGMN3LO,Ente17EMn4lo} have been regularized using
a different, non-local regulator of the form
\begin{equation} f_\Lambda^{\text{non-loc}}(p^2,p'^2)=\exp
\left[-(p^{2n}+p'{}^{2n})/\Lambda^{2n} \right].
\label{eq:nonlocal_reg}
\end{equation}

\begin{figure}[t]
\begin{center}
\includegraphics[width=0.48\textwidth,clip=]{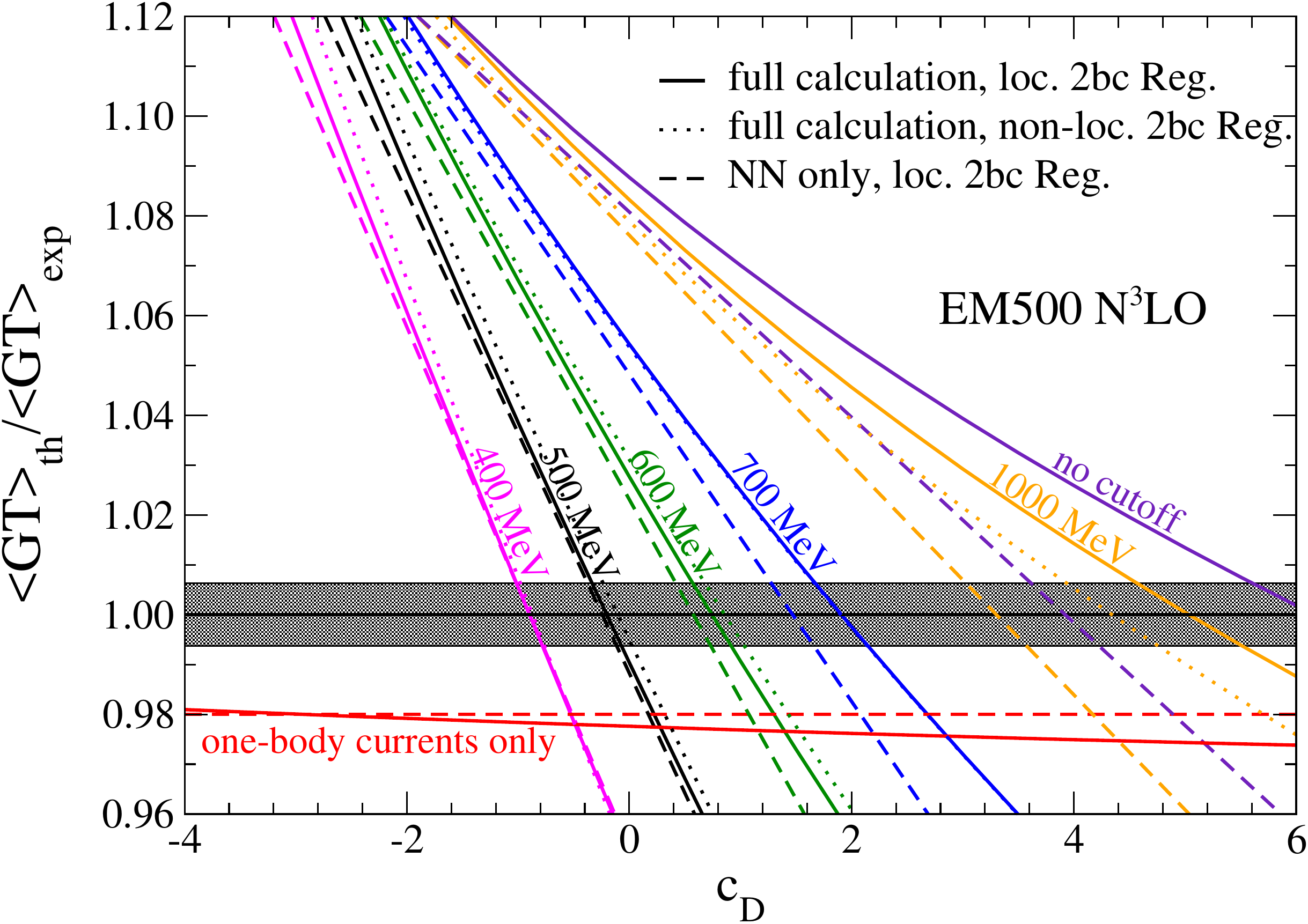}
\end{center}
\caption{(Color online) Ratio of calculated and experimental Gamow-Teller
matrix elements as a function of $c_D$ for different cutoff values
and regulators in the two-body currents, based on
the EM 500 MeV N$^3$LO potential of Ref.~\cite{Ente03EMN3LO}.
The solid (dotted) lines show results for nuclear states including 3N forces at
N$^2$LO using the $^3$H binding energy constraint of Fig.~\ref{fig:3Hfits} for
a local [non-local with $n = 2$, see Eq.(\ref{eq:nonlocal_reg})] regulator in the two-body currents. For comparison, we
also show results based on NN interactions only (dashed lines) and with 1b
currents only.  The width of the shaded band denotes the $2\sigma$
experimental uncertainty. }
\label{Fig_cD_Reg_dep}
\end{figure}

\begin{figure}[t]
\begin{center}
\includegraphics[width=0.48\textwidth,clip=]{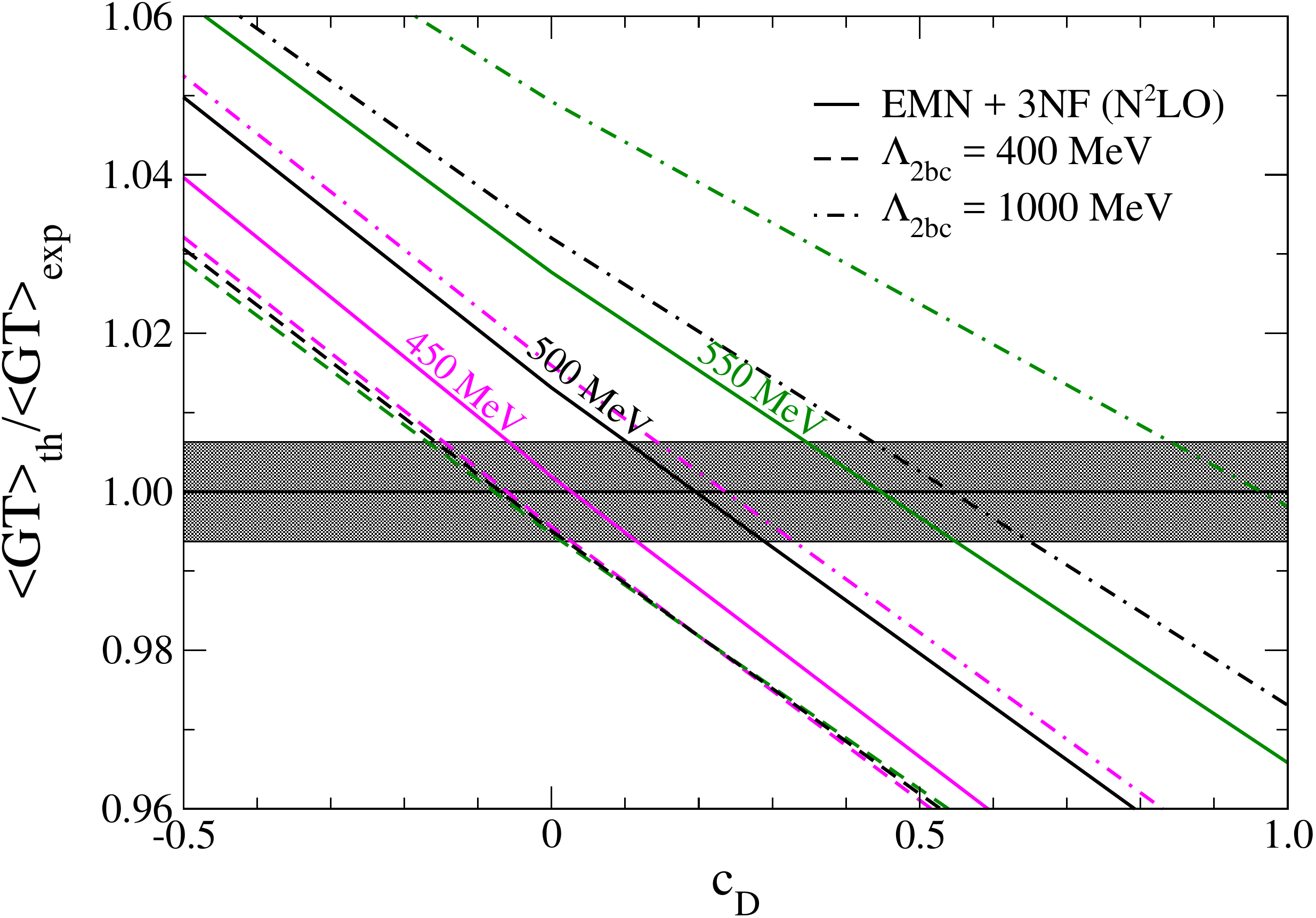}
\end{center}
\caption{(Color online) Ratio of calculated and experimental
Gamow-Teller matrix elements as a function of $c_D$ based on the EMN
potentials at order N$^2$LO of Ref.~\cite{Ente17EMn4lo}. We show results for the 2b current cutoff $\Lambda_\text{2bc}=400$~MeV (dashed lines) and $\Lambda_\text{2bc}=1000$~MeV (dash-dotted lines)
using a non-local regulator [see Eq.(\ref{eq:nonlocal_reg})] with
$n = 2$, whereas the solid lines show the cases 
when using the same cutoff values in the regulators for the interactions and currents.
The width of the shaded band denotes the $2\sigma$ experimental uncertainty.}
\label{Fig_cD_Reg_dep_EMN}
\end{figure}

{\it Results.} First we calculate the Fermi matrix element and obtain $\left<
\text{F}
\right>=0.9998$ for all potentials from Ref.~\cite{Ente03EMN3LO} independently
of the LECs in the 3N force, in good agreement with isospin conservation and
previous calculations~\cite{Schi98weakcap,Baro16beta}. We then focus on the
Gamow-Teller matrix element, which can be fitted to the $^3$H half-life using
Eq.~\eqref{eq:half-life}. We represent all terms of
Eq.~(\ref{Eq_axial_curr_q0}) in the partial-wave basis and evaluate the
reduced matrix element in Eq.~(\ref{eq:gt_me}).  Figure~\ref{Fig_cD_Reg_dep}
shows results for the Gamow-Teller matrix elements as a function of the $c_D$
values, evaluated for different cutoff values $\Lambda$ in the
2b current regulators of Eqs.~(\ref{eq:local_reg}) and (\ref{eq:nonlocal_reg}). 
We use no regulator for the 1b current, consistently with the calculation of $\left< \text{F}
\right>$.  Our results are presented in terms of the ratio of the calculated
and the experimental matrix element
$\langle\text{GT}\rangle_\text{exp}=\sqrt{3}\cdot 0.956$, which reproduces the
measured $^3$H half-life~\cite{Gazi08lec}. For each cutoff value, we show
results based on calculations including 3N forces (solid lines) and without 3N
forces (dashed lines) using for the 2b currents the local regulator of Eq.~(\ref{eq:local_reg}).
In addition we show results using the non-local regulator of
Eq.~(\ref{eq:nonlocal_reg}) in the 2b current. For comparison we also
show results including only 1b currents [see Eq.~(\ref{eq:GT_onebody})]. This
approximation underpredicts the experimental value by about 2\%, which
demonstrates the need for a small 2b-current contribution. For comparison,
Fig.~\ref{Fig_cD_Reg_dep_EMN} shows the corresponding Gamow-Teller matrix
elements as a function of the $c_D$ values for the different EMN potentials
plus 3N interactions at N$^2$LO. 

Figures~\ref{Fig_cD_Reg_dep} and \ref{Fig_cD_Reg_dep_EMN} show that for each
cutoff value the Gamow-Teller matrix elements exhibit a strong
$c_D$ dependence, which indicates the small degree of correlation between the $^3$H binding energy and lifetime and in principle allows for a precise determination of $c_D$
by fitting to the experimental $\langle\text{GT}\rangle_\text{exp}$ value.
However, the inferred $c_D$ values also depend sensitively on the regulator
applied to the current operator. While the cutoff $\Lambda=400$~MeV in
Fig.~\ref{Fig_cD_Reg_dep} yields a value of $c_D =-0.9$, a maximal value of
$c_D=6.0$ is found for unregularized current operators. In the latter case the
contributions of the current operator are cut off at high momenta solely by
the nuclear states. For $\Lambda=500$~MeV we reproduce the result of
Ref.~\cite{Gazi08lec} of $c_D=-0.24$ for a local regulator of the form
Eq.(\ref{eq:local_reg}). Note that for these calculations a non-local regulator
was used for the NN interactions and a local regulator for the currents.
Replacing the regulator in the 2b currents by a non-local one yields only
slight changes in the curves and consequently very similar $c_D$ values. 
In addition, Fig.~\ref{Fig_cD_Reg_dep} also shows that the sensitivity of the
Gamow-Teller matrix elements on 3N forces depends significantly on the cutoff
value. While for small cutoffs the fitted values for $c_D$ are to very good
approximation independent of 3N interactions (as argued in
Ref.~\cite{Gazi08lec}), for $\Lambda \gtrsim 600$ MeV the fits start to become
sensitive to contributions from 3N forces.

Figure~\ref{Fig_cD_Reg_dep_EMN} shows the corresponding results for the $c_D$
ranges based on the EMN NN interactions plus 3N interactions at N$^2$LO. The
solid lines show the results using the same cutoff values in the non-local
regulators for the currents and the NN and 3N interactions. The dashed and dash-dotted lines are
generated by changing the cutoff values for the currents to 400~MeV and 1000~MeV, respectively. The reduced sensitivity of the $c_D$ values on the cutoff
values compared to the results of Fig.~\ref{Fig_cD_Reg_dep} can be traced back
to the enhanced perturbativeness of the EMN potentials compared to the
previous EM 500 MeV potential~\cite{Hopp17WeinEVAn}. In particular, we found that contributions in
the three-body wave functions for the latter interaction extend to much higher
momenta and consequently a regulator has a stronger impact on the results for
small cutoff values.

These results highlight that the sensitivity of the Gamow-Teller matrix
elements on the regulators used for both forces and current operators can lead
to significant uncertainties for the extracted values of $c_D$. This sensitivity has also been studied in
Refs.~\cite{Marc12mucap,Cora14nmat,Samm15numat,Baro16beta}. Taking into
account the different regularization schemes used in these works the obtained $c_D$
variation is of similar size to the one we found for the EMN potentials shown in
Fig.~\ref{Fig_cD_Reg_dep_EMN}. Still, it is not obvious how a consistent choice
of regulators can be precisely defined in this context, especially when the
form of the employed regulator is different for forces and
currents, as is the case for the results using the local regulator shown in
Fig.~\ref{Fig_cD_Reg_dep}. Until these consistency constraints are taken into
account it is crucial to include the uncertainties due to the regulator
dependence when fitting LECs to the $\beta$ decay of $^3$H.

\begin{figure}[t]
\begin{center}
\includegraphics[width=0.48\textwidth,clip=]{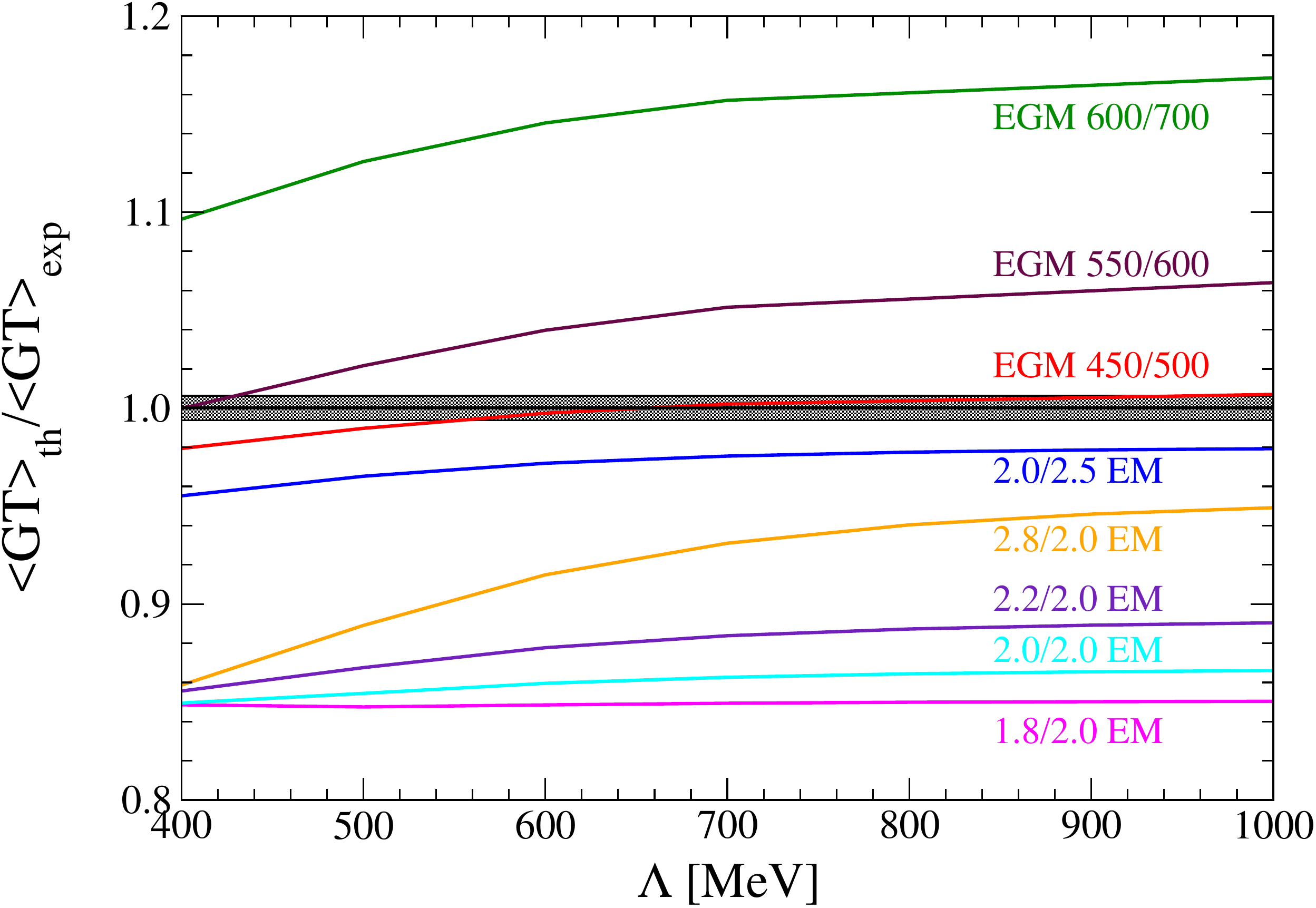}
\end{center}
\caption{(Color online) Ratio of calculated and experimental Gamow-Teller
matrix elements as a function of the cutoff $\Lambda$ in the non-local
regulator [$n=2$, see Eq.~(\ref{eq:nonlocal_reg})] for a set of chiral
interactions, using different fitting observables: the results labeled 'EM'
are based on NN plus 3N interactions, for which the $c_D$ and $c_E$ values are
fit to the binding energy of $^3$H and the charge radius of $^4$He (see
Ref.~\cite{Hebe11fits} for details). The results labeled 'EGM' are based on NN
plus 3N forces fitted to the binding energy of $^3$H and the neutron-deuteron
scattering length~\cite{Epel06PPNP}. The width of the shaded band denotes the
$2\sigma$ experimental uncertainty.}
\label{fig:SRG_fits}
\end{figure}

Given the different possible observables used to determine $c_D$ and $c_E$ the
question arises to what extent these fits are compatible with extractions from
the $^3$H $\beta$ decay. Figure~\ref{fig:SRG_fits} shows the results for the
Gamow-Teller matrix elements as a function of the 2b current cutoff for
interactions fitted to the binding energy of $^3$H and the charge radius of
$^4$He~\cite{Hebe11fits} (denoted as EM), as well as interactions fitted to
the binding energy of $^3$H and the neutron-deuteron scattering
length~\cite{Epel02fewbody} (EGM). The different EGM interactions are labeled
by the value of the internal resolution scales in MeV (see
Ref.~\cite{Epel06PPNP} for details), whereas the EM interactions are labeled
by the similarity-renormalization-group resolution scale in units of fm$^{-1}$
(see Ref.~\cite{Hebe11fits} for details). Again, we find that the sensitivity
of the results on the cutoff $\Lambda$ tends to become stronger with
increasing resolution scales and reduced perturbativeness. Second, in
general it is not possible to reproduce the experimental value of the $^3$H
$\beta$-decay Gamow-Teller matrix element based on LECs extracted from fits to
other observables, even when the cutoff dependence of the results is taken
into account. This discrepancy might be caused by the application of chiral forces
and currents at different chiral orders as suggested in
Ref.~\cite{Baro16beta} and also suggests that the uncertainty of the $^3$H $\beta$ decay half-life is larger than the effect from 2b currents, especially at N$^2$LO.

We also note that fitting the Gamow-Teller matrix element within
experimental uncertainties neglects the fact that the calculations
are constrained to a particular chiral order, which can be associated
with an uncertainty due to the truncation in the chiral expansion 
(e.g., for recent work, see Refs.~\cite{Epel15improved,Bind15Fewbody}). This illustrates, in agreement
with the main result of this work, that previous works have 
underestimated the uncertainties when fitting LECs to the triton half-life.
We leave such consistent order by order calculations to future work.

\begin{figure}[t]
\begin{center}
\includegraphics[width=0.48\textwidth,clip=]{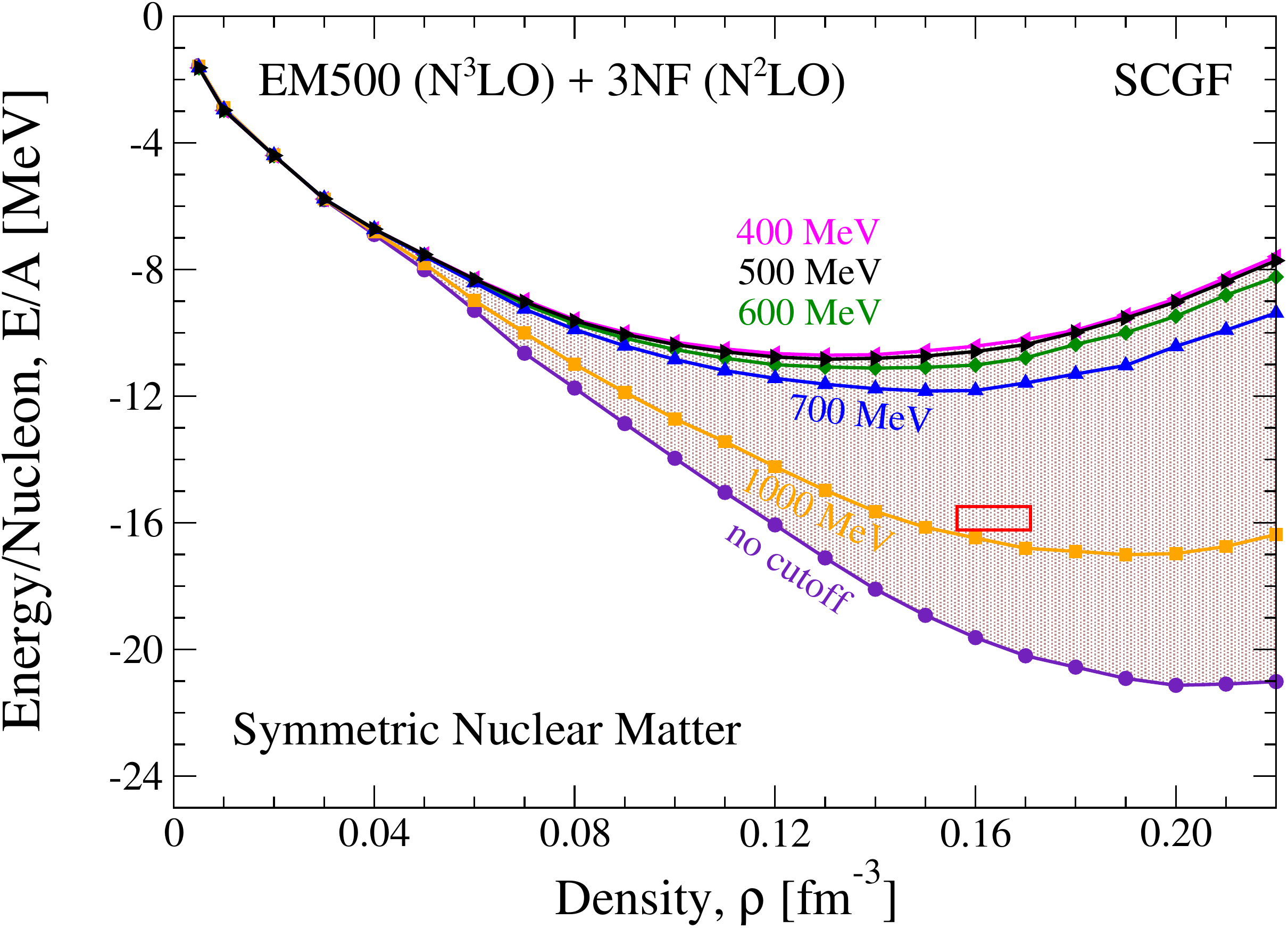}
\end{center}
\caption{(Color online) Energy per nucleon of symmetric nuclear matter as a
function of nucleon density obtained within the self-consistent Green's
function approach~\cite{Carb14SCGFdd}. Results are based on the NN EM 500 MeV
at N$^3$LO, including normal-ordered 3N interaction contributions at N$^2$LO.
The curves correspond to different $c_D$ and $c_E$ values obtained according
to Figs.~\ref{fig:3Hfits} and~\ref{Fig_cD_Reg_dep}. The box describes the
range for the empirical saturation point provided by mean-field
calculations~\cite{Dutra12skyrme}.}
\label{Fig_snm_ener}
\end{figure}

\begin{figure}[t]
\begin{center}
\includegraphics[width=0.48\textwidth,clip=]{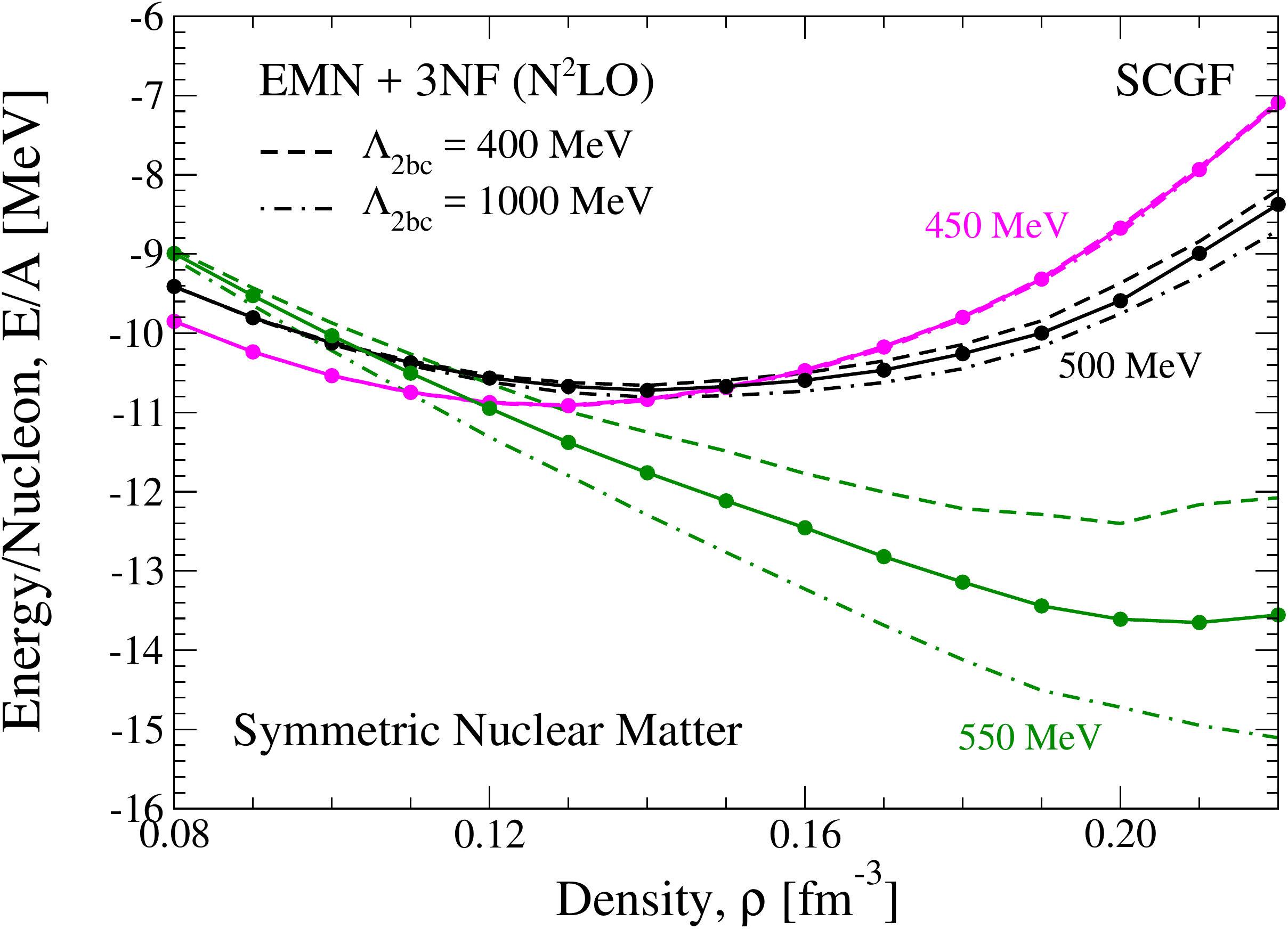}
\end{center}
\caption{(Color online) Same as Fig.~\ref{Fig_snm_ener} but
showing results based on the EMN potentials at N$^2$LO. The curves correspond to
different $c_D$ and $c_E$ values obtained according to Figs.~\ref{fig:3Hfits}
and~\ref{Fig_cD_Reg_dep_EMN}.
The different lines correspond to 2b currents cutoffs equivalent to the interaction (solid), $\Lambda_\text{2bc}=400$~MeV (dashed), and $\Lambda_\text{2bc}=1000$~MeV (dash-dotted) as  shown in Fig.~\ref{Fig_cD_Reg_dep_EMN}.}
\label{Fig_snm_ener_emn}
\end{figure}

Based on the obtained uncertainty ranges for $c_D$ and $c_E$ we also explore
the resulting uncertainties for many-body observables by calculating the
energy per nucleon of symmetric nuclear matter within the self-consistent
Green's function framework~\cite{Carb14SCGFdd}. In this nonperturbative many-body approach 3N
forces are included via a normal-ordering procedure with respect to the fully
correlated reference state~\cite{Carb13SCGF3B,Carb14SCGFdd}.
Figures~\ref{Fig_snm_ener} and \ref{Fig_snm_ener_emn} show the energy per
nucleon $E/A$ of symmetric nuclear matter as a function of the density $\rho$
using the NN and 3N interactions used in Figs.~\ref{Fig_cD_Reg_dep} and
\ref{Fig_cD_Reg_dep_EMN}, respectively. For the EM 500 MeV potential we find a
strong dependence on the $c_D$ and $c_E$ values corresponding to different
cutoff values in the local regulator of the 2b currents. The saturation energy
ranges from $E/A\sim-11$ MeV, for smaller cutoffs, to $E/A\sim-21$ MeV, for
the LECs corresponding to the unregularized 2b current.  For cutoffs in the
range $\Lambda=400-700$ MeV, the saturation density takes the values
$\rho\sim0.13-0.15$ fm$^{-3}$. Performing corresponding calculations using the
EMN potentials we find a smaller dependence on the cutoff
values. In Fig.~\ref{Fig_snm_ener_emn} we show results for different values of $c_D$ and $c_E$ as obtained from the fit to the Gamow-Teller matrix element in Fig.~\ref{Fig_cD_Reg_dep_EMN}. First, we notice that the variation with respect to change in the 2b current cutoff grows with increasing cutoff value of the interaction. This is related to the increasing range of $c_D$ values obtained in Fig.~\ref{Fig_cD_Reg_dep_EMN}. Furthermore, the variation due to the change of the 2b current cutoff of the 500~MeV EMN potential is much smaller than in Fig.~\ref{Fig_snm_ener} for the EM 500~MeV potential. As mentioned before this can be explained by the enhanced perturbativeness of the EMN potentials. However, as the potentials become less perturbative, which is the case for the $\Lambda=550$~MeV potential, the variation increases highlighting the relevance of the 2b current cutoff.  
Finally, we state that for all EMN potentials it is not possible to reproduce the saturation point based on the fits to the $^3$H $\beta$ decay.

These results indicate that the uncertainties of the LECs due to the regulator
dependence of the nuclear forces and currents can lead to significant
uncertainties for nuclear structure observables. The observed dependence may
also be related to possible inconsistencies in the power counting in the currents due to
the non-trivial enhancement of some contributions~\cite{Vald15pcCurr}. In
order to systematically reduce these uncertainties more detailed studies are
required to find a consistent way of regularizing nuclear interactions and
currents ensuring also the continuity equation. In Ref.~\cite{Kreb16axial} it
was shown that the currents and interactions indeed fulfill the continuity
equation at the operator level, i.e., for infinite cutoffs. Generalizing this
analysis to regularized matrix elements will provide additional nontrivial
constraints for a consistent way of regularizing electroweak currents.

{\it Summary.} We have studied the uncertainties in constraining the LECs of
the leading-order 3N interactions, $c_D$ and $c_E$, via fits to $^3$H $\beta$
decay. We found that the extracted values of $c_D$ generally exhibit a
significant dependence on the cutoff scale used for the currents,
whereas the degree of uncertainty is correlated to the perturbativeness of the
employed nuclear interactions. These additional uncertainties need to be taken
into account when including 3N interactions based on such fits in nuclear
structure calculations, and can be sizable as illustrated by nuclear matter
calculations. Furthermore, we analyzed the Gamow-Teller matrix elements
calculated with different chiral forces and $c_D$ and $c_E$ values obtained
through fits to other observables. We found that it is generally not possible
to simultaneously fit all experimental observables even when the uncertainty
due to sensitivity to the regulator scheme is taken into account. Our studies
indicate that the importance of the regularization scheme and scale in chiral
many-body currents needs to be studied more carefully than in previous works, and can lead
to additional uncertainties in the many-body calculations. Further studies need
to be performed in order to work out a consistent way of regularizing chiral
forces and currents.

{\it Acknowledgments.} We thank R.\ J.\ Furnstahl, D.\ Gazit and K.\ A.\ Wendt
for useful discussions. This work was supported by the DFG through Grant SFB
1245, the ERC Grant No.~307986 STRONGINT, and the JSPS Grant-in-Aid for
Scientific Research No.\ 26$\cdot$04323. JM\ was also supported by an
International Research Fellowship from JSPS. AC acknowledges support by the Alexander von Humboldt Foundation through a Humboldt Research Fellowship for Postdoctoral Researchers.

\bibliographystyle{apsrev}
\bibliography{strongint.bib}

\begin{thebibliography}{65}
\expandafter\ifx\csname natexlab\endcsname\relax\def\natexlab#1{#1}\fi
\expandafter\ifx\csname bibnamefont\endcsname\relax
  \def\bibnamefont#1{#1}\fi
\expandafter\ifx\csname bibfnamefont\endcsname\relax
  \def\bibfnamefont#1{#1}\fi
\expandafter\ifx\csname citenamefont\endcsname\relax
  \def\citenamefont#1{#1}\fi
\expandafter\ifx\csname url\endcsname\relax
  \def\url#1{\texttt{#1}}\fi
\expandafter\ifx\csname urlprefix\endcsname\relax\def\urlprefix{URL }\fi
\providecommand{\bibinfo}[2]{#2}
\providecommand{\eprint}[2][]{\url{#2}}

\bibitem[{\citenamefont{Gezerlis et~al.}(2013)\citenamefont{Gezerlis, Tews,
  Epelbaum, Gandolfi, Hebeler, Nogga, and Schwenk}}]{Geze13QMCchi}
\bibinfo{author}{\bibfnamefont{A.}~\bibnamefont{Gezerlis}},
  \bibinfo{author}{\bibfnamefont{I.}~\bibnamefont{Tews}},
  \bibinfo{author}{\bibfnamefont{E.}~\bibnamefont{Epelbaum}},
  \bibinfo{author}{\bibfnamefont{S.}~\bibnamefont{Gandolfi}},
  \bibinfo{author}{\bibfnamefont{K.}~\bibnamefont{Hebeler}},
  \bibinfo{author}{\bibfnamefont{A.}~\bibnamefont{Nogga}}, \bibnamefont{and}
  \bibinfo{author}{\bibfnamefont{A.}~\bibnamefont{Schwenk}},
  \bibinfo{journal}{Phys. Rev. Lett.} \textbf{\bibinfo{volume}{111}},
  \bibinfo{pages}{032501} (\bibinfo{year}{2013}).

\bibitem[{\citenamefont{Ekstr\"om et~al.}(2015)\citenamefont{Ekstr\"om, Jansen,
  Wendt, Hagen, Papenbrock, Carlsson, Forss\'en, Hjorth-Jensen, Navr\'atil, and
  Nazarewicz}}]{Ekst15sat}
\bibinfo{author}{\bibfnamefont{A.}~\bibnamefont{Ekstr\"om}},
  \bibinfo{author}{\bibfnamefont{G.~R.} \bibnamefont{Jansen}},
  \bibinfo{author}{\bibfnamefont{K.~A.} \bibnamefont{Wendt}},
  \bibinfo{author}{\bibfnamefont{G.}~\bibnamefont{Hagen}},
  \bibinfo{author}{\bibfnamefont{T.}~\bibnamefont{Papenbrock}},
  \bibinfo{author}{\bibfnamefont{B.~D.} \bibnamefont{Carlsson}},
  \bibinfo{author}{\bibfnamefont{C.}~\bibnamefont{Forss\'en}},
  \bibinfo{author}{\bibfnamefont{M.}~\bibnamefont{Hjorth-Jensen}},
  \bibinfo{author}{\bibfnamefont{P.}~\bibnamefont{Navr\'atil}},
  \bibnamefont{and}
  \bibinfo{author}{\bibfnamefont{W.}~\bibnamefont{Nazarewicz}},
  \bibinfo{journal}{Phys. Rev. C} \textbf{\bibinfo{volume}{91}},
  \bibinfo{pages}{051301(R)} (\bibinfo{year}{2015}).

\bibitem[{\citenamefont{Epelbaum
  et~al.}(2015{\natexlab{a}})\citenamefont{Epelbaum, Krebs, and
  Mei{\ss}ner}}]{Epel15NNn4lo}
\bibinfo{author}{\bibfnamefont{E.}~\bibnamefont{Epelbaum}},
  \bibinfo{author}{\bibfnamefont{H.}~\bibnamefont{Krebs}}, \bibnamefont{and}
  \bibinfo{author}{\bibfnamefont{U.-G.} \bibnamefont{Mei{\ss}ner}},
  \bibinfo{journal}{Phys. Rev. Lett.} \textbf{\bibinfo{volume}{115}},
  \bibinfo{pages}{122301} (\bibinfo{year}{2015}{\natexlab{a}}).

\bibitem[{\citenamefont{Carlsson et~al.}(2016)\citenamefont{Carlsson,
  Ekstr{\"o}m, Forss\'en, Str{\"o}mberg, Jansen, Lilja, Lindby, Mattsson, and
  Wendt}}]{Carl15sim}
\bibinfo{author}{\bibfnamefont{B.~D.} \bibnamefont{Carlsson}},
  \bibinfo{author}{\bibfnamefont{A.}~\bibnamefont{Ekstr{\"o}m}},
  \bibinfo{author}{\bibfnamefont{C.}~\bibnamefont{Forss\'en}},
  \bibinfo{author}{\bibfnamefont{D.~F.} \bibnamefont{Str{\"o}mberg}},
  \bibinfo{author}{\bibfnamefont{G.~R.} \bibnamefont{Jansen}},
  \bibinfo{author}{\bibfnamefont{O.}~\bibnamefont{Lilja}},
  \bibinfo{author}{\bibfnamefont{M.}~\bibnamefont{Lindby}},
  \bibinfo{author}{\bibfnamefont{B.~A.} \bibnamefont{Mattsson}},
  \bibnamefont{and} \bibinfo{author}{\bibfnamefont{K.~A.} \bibnamefont{Wendt}},
  \bibinfo{journal}{Phys. Rev. X} \textbf{\bibinfo{volume}{6}},
  \bibinfo{pages}{011019} (\bibinfo{year}{2016}).

\bibitem[{\citenamefont{Epelbaum et~al.}(2009)\citenamefont{Epelbaum, Hammer,
  and Mei{\ss}ner}}]{Epel09RMP}
\bibinfo{author}{\bibfnamefont{E.}~\bibnamefont{Epelbaum}},
  \bibinfo{author}{\bibfnamefont{H.-W.} \bibnamefont{Hammer}},
  \bibnamefont{and} \bibinfo{author}{\bibfnamefont{U.-G.}
  \bibnamefont{Mei{\ss}ner}}, \bibinfo{journal}{Rev. Mod. Phys.}
  \textbf{\bibinfo{volume}{81}}, \bibinfo{pages}{1773} (\bibinfo{year}{2009}).

\bibitem[{\citenamefont{Machleidt and Entem}(2011)}]{Mach11PR}
\bibinfo{author}{\bibfnamefont{R.}~\bibnamefont{Machleidt}} \bibnamefont{and}
  \bibinfo{author}{\bibfnamefont{D.~R.} \bibnamefont{Entem}},
  \bibinfo{journal}{Phys. Rep.} \textbf{\bibinfo{volume}{503}},
  \bibinfo{pages}{1} (\bibinfo{year}{2011}).

\bibitem[{\citenamefont{Hammer et~al.}(2013)\citenamefont{Hammer, Nogga, and
  Schwenk}}]{Hamm13RMP}
\bibinfo{author}{\bibfnamefont{H.-W.} \bibnamefont{Hammer}},
  \bibinfo{author}{\bibfnamefont{A.}~\bibnamefont{Nogga}}, \bibnamefont{and}
  \bibinfo{author}{\bibfnamefont{A.}~\bibnamefont{Schwenk}},
  \bibinfo{journal}{Rev. Mod. Phys.} \textbf{\bibinfo{volume}{85}},
  \bibinfo{pages}{197} (\bibinfo{year}{2013}).

\bibitem[{\citenamefont{Bacca and Pastore}(2014)}]{Bacca14JPG}
\bibinfo{author}{\bibfnamefont{S.}~\bibnamefont{Bacca}} \bibnamefont{and}
  \bibinfo{author}{\bibfnamefont{S.}~\bibnamefont{Pastore}},
  \bibinfo{journal}{J. Phys. G} \textbf{\bibinfo{volume}{41}},
  \bibinfo{pages}{123002} (\bibinfo{year}{2014}).

\bibitem[{\citenamefont{Weinberg}(1991)}]{Wein91chNp}
\bibinfo{author}{\bibfnamefont{S.}~\bibnamefont{Weinberg}},
  \bibinfo{journal}{Nucl. Phys. B} \textbf{\bibinfo{volume}{363}},
  \bibinfo{pages}{3} (\bibinfo{year}{1991}).

\bibitem[{\citenamefont{Entem et~al.}(2015)\citenamefont{Entem, Kaiser,
  Machleidt, and Nosyk}}]{Ente14NNn4lo}
\bibinfo{author}{\bibfnamefont{D.~R.} \bibnamefont{Entem}},
  \bibinfo{author}{\bibfnamefont{N.}~\bibnamefont{Kaiser}},
  \bibinfo{author}{\bibfnamefont{R.}~\bibnamefont{Machleidt}},
  \bibnamefont{and} \bibinfo{author}{\bibfnamefont{Y.}~\bibnamefont{Nosyk}},
  \bibinfo{journal}{Phys. Rev. C} \textbf{\bibinfo{volume}{91}},
  \bibinfo{pages}{014002} (\bibinfo{year}{2015}).

\bibitem[{\citenamefont{Ishikawa and Robilotta}(2007)}]{Ishi073N2p}
\bibinfo{author}{\bibfnamefont{S.}~\bibnamefont{Ishikawa}} \bibnamefont{and}
  \bibinfo{author}{\bibfnamefont{M.~R.} \bibnamefont{Robilotta}},
  \bibinfo{journal}{Phys. Rev. C} \textbf{\bibinfo{volume}{76}},
  \bibinfo{pages}{014006} (\bibinfo{year}{2007}).

\bibitem[{\citenamefont{Bernard et~al.}(2008)\citenamefont{Bernard, Epelbaum,
  Krebs, and Mei{\ss}ner}}]{Bern083Nlong}
\bibinfo{author}{\bibfnamefont{V.}~\bibnamefont{Bernard}},
  \bibinfo{author}{\bibfnamefont{E.}~\bibnamefont{Epelbaum}},
  \bibinfo{author}{\bibfnamefont{H.}~\bibnamefont{Krebs}}, \bibnamefont{and}
  \bibinfo{author}{\bibfnamefont{U.-G.} \bibnamefont{Mei{\ss}ner}},
  \bibinfo{journal}{Phys. Rev. C} \textbf{\bibinfo{volume}{77}},
  \bibinfo{pages}{064004} (\bibinfo{year}{2008}).

\bibitem[{\citenamefont{Bernard et~al.}(2011)\citenamefont{Bernard, Epelbaum,
  Krebs, and Mei{\ss}ner}}]{Bern113Nshort}
\bibinfo{author}{\bibfnamefont{V.}~\bibnamefont{Bernard}},
  \bibinfo{author}{\bibfnamefont{E.}~\bibnamefont{Epelbaum}},
  \bibinfo{author}{\bibfnamefont{H.}~\bibnamefont{Krebs}}, \bibnamefont{and}
  \bibinfo{author}{\bibfnamefont{U.-G.} \bibnamefont{Mei{\ss}ner}},
  \bibinfo{journal}{Phys. Rev. C} \textbf{\bibinfo{volume}{84}},
  \bibinfo{pages}{054001} (\bibinfo{year}{2011}).

\bibitem[{\citenamefont{Epelbaum}(2006{\natexlab{a}})}]{Epel064N}
\bibinfo{author}{\bibfnamefont{E.}~\bibnamefont{Epelbaum}},
  \bibinfo{journal}{Phys. Lett. B} \textbf{\bibinfo{volume}{639}},
  \bibinfo{pages}{456} (\bibinfo{year}{2006}{\natexlab{a}}).

\bibitem[{\citenamefont{Krebs et~al.}(2017)\citenamefont{Krebs, Epelbaum, and
  Meißner}}]{Kreb16axial}
\bibinfo{author}{\bibfnamefont{H.}~\bibnamefont{Krebs}},
  \bibinfo{author}{\bibfnamefont{E.}~\bibnamefont{Epelbaum}}, \bibnamefont{and}
  \bibinfo{author}{\bibfnamefont{U.~G.} \bibnamefont{Meißner}},
  \bibinfo{journal}{Annals Phys.} \textbf{\bibinfo{volume}{378}},
  \bibinfo{pages}{317} (\bibinfo{year}{2017}).

\bibitem[{\citenamefont{Baroni et~al.}(2016{\natexlab{a}})\citenamefont{Baroni,
  Girlanda, Pastore, Schiavilla, and Viviani}}]{Baro15axial}
\bibinfo{author}{\bibfnamefont{A.}~\bibnamefont{Baroni}},
  \bibinfo{author}{\bibfnamefont{L.}~\bibnamefont{Girlanda}},
  \bibinfo{author}{\bibfnamefont{S.}~\bibnamefont{Pastore}},
  \bibinfo{author}{\bibfnamefont{R.}~\bibnamefont{Schiavilla}},
  \bibnamefont{and} \bibinfo{author}{\bibfnamefont{M.}~\bibnamefont{Viviani}},
  \bibinfo{journal}{Phys. Rev. C} \textbf{\bibinfo{volume}{93}},
  \bibinfo{pages}{015501} (\bibinfo{year}{2016}{\natexlab{a}}).

\bibitem[{\citenamefont{K{\"o}lling et~al.}(2009)\citenamefont{K{\"o}lling,
  Epelbaum, Krebs, and Mei{\ss}ner}}]{Koll09curr}
\bibinfo{author}{\bibfnamefont{S.}~\bibnamefont{K{\"o}lling}},
  \bibinfo{author}{\bibfnamefont{E.}~\bibnamefont{Epelbaum}},
  \bibinfo{author}{\bibfnamefont{H.}~\bibnamefont{Krebs}}, \bibnamefont{and}
  \bibinfo{author}{\bibfnamefont{U.-G.} \bibnamefont{Mei{\ss}ner}},
  \bibinfo{journal}{Phys. Rev. C} \textbf{\bibinfo{volume}{80}},
  \bibinfo{pages}{045502} (\bibinfo{year}{2009}).

\bibitem[{\citenamefont{K\"olling et~al.}(2011)\citenamefont{K\"olling,
  Epelbaum, Krebs, and Mei\ss{}ner}}]{Koll11em2b}
\bibinfo{author}{\bibfnamefont{S.}~\bibnamefont{K\"olling}},
  \bibinfo{author}{\bibfnamefont{E.}~\bibnamefont{Epelbaum}},
  \bibinfo{author}{\bibfnamefont{H.}~\bibnamefont{Krebs}}, \bibnamefont{and}
  \bibinfo{author}{\bibfnamefont{U.-G.} \bibnamefont{Mei\ss{}ner}},
  \bibinfo{journal}{Phys. Rev. C} \textbf{\bibinfo{volume}{84}},
  \bibinfo{pages}{054008} (\bibinfo{year}{2011}).

\bibitem[{\citenamefont{Pastore et~al.}(2009)\citenamefont{Pastore, Girlanda,
  Schiavilla, Viviani, and Wiringa}}]{Past09EMcurr}
\bibinfo{author}{\bibfnamefont{S.}~\bibnamefont{Pastore}},
  \bibinfo{author}{\bibfnamefont{L.}~\bibnamefont{Girlanda}},
  \bibinfo{author}{\bibfnamefont{R.}~\bibnamefont{Schiavilla}},
  \bibinfo{author}{\bibfnamefont{M.}~\bibnamefont{Viviani}}, \bibnamefont{and}
  \bibinfo{author}{\bibfnamefont{R.~B.} \bibnamefont{Wiringa}},
  \bibinfo{journal}{Phys. Rev. C} \textbf{\bibinfo{volume}{80}},
  \bibinfo{pages}{034004} (\bibinfo{year}{2009}).

\bibitem[{\citenamefont{Pastore et~al.}(2011)\citenamefont{Pastore, Girlanda,
  Schiavilla, and Viviani}}]{Past11EMcharge}
\bibinfo{author}{\bibfnamefont{S.}~\bibnamefont{Pastore}},
  \bibinfo{author}{\bibfnamefont{L.}~\bibnamefont{Girlanda}},
  \bibinfo{author}{\bibfnamefont{R.}~\bibnamefont{Schiavilla}},
  \bibnamefont{and} \bibinfo{author}{\bibfnamefont{M.}~\bibnamefont{Viviani}},
  \bibinfo{journal}{Phys. Rev. C} \textbf{\bibinfo{volume}{84}},
  \bibinfo{pages}{024001} (\bibinfo{year}{2011}).

\bibitem[{\citenamefont{Piarulli et~al.}(2013)\citenamefont{Piarulli, Girlanda,
  Marcucci, Pastore, Schiavilla, and Viviani}}]{Piar13EMcurr}
\bibinfo{author}{\bibfnamefont{M.}~\bibnamefont{Piarulli}},
  \bibinfo{author}{\bibfnamefont{L.}~\bibnamefont{Girlanda}},
  \bibinfo{author}{\bibfnamefont{L.~E.} \bibnamefont{Marcucci}},
  \bibinfo{author}{\bibfnamefont{S.}~\bibnamefont{Pastore}},
  \bibinfo{author}{\bibfnamefont{R.}~\bibnamefont{Schiavilla}},
  \bibnamefont{and} \bibinfo{author}{\bibfnamefont{M.}~\bibnamefont{Viviani}},
  \bibinfo{journal}{Phys. Rev. C} \textbf{\bibinfo{volume}{87}},
  \bibinfo{pages}{014006} (\bibinfo{year}{2013}).

\bibitem[{\citenamefont{Gazit et~al.}(2009)\citenamefont{Gazit, Quaglioni, and
  Navr{\'a}til}}]{Gazi08lec}
\bibinfo{author}{\bibfnamefont{D.}~\bibnamefont{Gazit}},
  \bibinfo{author}{\bibfnamefont{S.}~\bibnamefont{Quaglioni}},
  \bibnamefont{and}
  \bibinfo{author}{\bibfnamefont{P.}~\bibnamefont{Navr{\'a}til}},
  \bibinfo{journal}{Phys. Rev. Lett.} \textbf{\bibinfo{volume}{103}},
  \bibinfo{pages}{102502} (\bibinfo{year}{2009}).

\bibitem[{\citenamefont{Marcucci et~al.}(2012)\citenamefont{Marcucci, Kievsky,
  Rosati, Schiavilla, and Viviani}}]{Marc12mucap}
\bibinfo{author}{\bibfnamefont{L.~E.} \bibnamefont{Marcucci}},
  \bibinfo{author}{\bibfnamefont{A.}~\bibnamefont{Kievsky}},
  \bibinfo{author}{\bibfnamefont{S.}~\bibnamefont{Rosati}},
  \bibinfo{author}{\bibfnamefont{R.}~\bibnamefont{Schiavilla}},
  \bibnamefont{and} \bibinfo{author}{\bibfnamefont{M.}~\bibnamefont{Viviani}},
  \bibinfo{journal}{Phys. Rev. Lett.} \textbf{\bibinfo{volume}{108}},
  \bibinfo{pages}{052502} (\bibinfo{year}{2012}).

\bibitem[{\citenamefont{Ekstr{\"o}m et~al.}(2014)\citenamefont{Ekstr{\"o}m,
  Jansen, Wendt, Hagen, Papenbrock et~al.}}]{Ekst14GT2bc}
\bibinfo{author}{\bibfnamefont{A.}~\bibnamefont{Ekstr{\"o}m}},
  \bibinfo{author}{\bibfnamefont{G.~R.} \bibnamefont{Jansen}},
  \bibinfo{author}{\bibfnamefont{K.~A.} \bibnamefont{Wendt}},
  \bibinfo{author}{\bibfnamefont{G.}~\bibnamefont{Hagen}},
  \bibinfo{author}{\bibfnamefont{T.}~\bibnamefont{Papenbrock}},
  \bibnamefont{et~al.}, \bibinfo{journal}{Phys. Rev. Lett.}
  \textbf{\bibinfo{volume}{113}}, \bibinfo{pages}{262504}
  (\bibinfo{year}{2014}).

\bibitem[{\citenamefont{Baroni et~al.}(2016{\natexlab{b}})\citenamefont{Baroni,
  Girlanda, Kievsky, Marcucci, Schiavilla, and Viviani}}]{Baro16beta}
\bibinfo{author}{\bibfnamefont{A.}~\bibnamefont{Baroni}},
  \bibinfo{author}{\bibfnamefont{L.}~\bibnamefont{Girlanda}},
  \bibinfo{author}{\bibfnamefont{A.}~\bibnamefont{Kievsky}},
  \bibinfo{author}{\bibfnamefont{L.~E.} \bibnamefont{Marcucci}},
  \bibinfo{author}{\bibfnamefont{R.}~\bibnamefont{Schiavilla}},
  \bibnamefont{and} \bibinfo{author}{\bibfnamefont{M.}~\bibnamefont{Viviani}},
  \bibinfo{journal}{Phys. Rev. C} \textbf{\bibinfo{volume}{94}},
  \bibinfo{pages}{024003} (\bibinfo{year}{2016}{\natexlab{b}}).

\bibitem[{\citenamefont{Hebeler et~al.}(2011)\citenamefont{Hebeler, Bogner,
  Furnstahl, Nogga, and Schwenk}}]{Hebe11fits}
\bibinfo{author}{\bibfnamefont{K.}~\bibnamefont{Hebeler}},
  \bibinfo{author}{\bibfnamefont{S.~K.} \bibnamefont{Bogner}},
  \bibinfo{author}{\bibfnamefont{R.~J.} \bibnamefont{Furnstahl}},
  \bibinfo{author}{\bibfnamefont{A.}~\bibnamefont{Nogga}}, \bibnamefont{and}
  \bibinfo{author}{\bibfnamefont{A.}~\bibnamefont{Schwenk}},
  \bibinfo{journal}{Phys. Rev. C} \textbf{\bibinfo{volume}{83}},
  \bibinfo{pages}{031301(R)} (\bibinfo{year}{2011}).

\bibitem[{\citenamefont{Epelbaum et~al.}(2002)\citenamefont{Epelbaum, Nogga,
  Gl{\"o}ckle, Kamada, Mei{\ss}ner, and Wita{\l}a}}]{Epel02fewbody}
\bibinfo{author}{\bibfnamefont{E.}~\bibnamefont{Epelbaum}},
  \bibinfo{author}{\bibfnamefont{A.}~\bibnamefont{Nogga}},
  \bibinfo{author}{\bibfnamefont{W.}~\bibnamefont{Gl{\"o}ckle}},
  \bibinfo{author}{\bibfnamefont{H.}~\bibnamefont{Kamada}},
  \bibinfo{author}{\bibfnamefont{U.-G.} \bibnamefont{Mei{\ss}ner}},
  \bibnamefont{and}
  \bibinfo{author}{\bibfnamefont{H.}~\bibnamefont{Wita{\l}a}},
  \bibinfo{journal}{Phys. Rev. C} \textbf{\bibinfo{volume}{66}},
  \bibinfo{pages}{064001} (\bibinfo{year}{2002}).

\bibitem[{\citenamefont{Lynn et~al.}(2016)\citenamefont{Lynn, Tews, Carlson,
  Gandolfi, Gezerlis, Schmidt, and Schwenk}}]{Lynn16QMC3N}
\bibinfo{author}{\bibfnamefont{J.~E.} \bibnamefont{Lynn}},
  \bibinfo{author}{\bibfnamefont{I.}~\bibnamefont{Tews}},
  \bibinfo{author}{\bibfnamefont{J.}~\bibnamefont{Carlson}},
  \bibinfo{author}{\bibfnamefont{S.}~\bibnamefont{Gandolfi}},
  \bibinfo{author}{\bibfnamefont{A.}~\bibnamefont{Gezerlis}},
  \bibinfo{author}{\bibfnamefont{K.~E.} \bibnamefont{Schmidt}},
  \bibnamefont{and} \bibinfo{author}{\bibfnamefont{A.}~\bibnamefont{Schwenk}},
  \bibinfo{journal}{Phys. Rev. Lett.} \textbf{\bibinfo{volume}{116}},
  \bibinfo{pages}{062501} (\bibinfo{year}{2016}).

\bibitem[{\citenamefont{Gardestig and Phillips}(2006)}]{Gard06weak3nf}
\bibinfo{author}{\bibfnamefont{A.}~\bibnamefont{Gardestig}} \bibnamefont{and}
  \bibinfo{author}{\bibfnamefont{D.~R.} \bibnamefont{Phillips}},
  \bibinfo{journal}{Phys. Rev. Lett.} \textbf{\bibinfo{volume}{96}},
  \bibinfo{pages}{232301} (\bibinfo{year}{2006}).

\bibitem[{\citenamefont{Barrett et~al.}(2013)\citenamefont{Barrett,
  Navr{\'a}til, and Vary}}]{Barr13PPNP}
\bibinfo{author}{\bibfnamefont{B.~R.} \bibnamefont{Barrett}},
  \bibinfo{author}{\bibfnamefont{P.}~\bibnamefont{Navr{\'a}til}},
  \bibnamefont{and} \bibinfo{author}{\bibfnamefont{J.~P.} \bibnamefont{Vary}},
  \bibinfo{journal}{Prog. Part. Nucl. Phys.} \textbf{\bibinfo{volume}{69}},
  \bibinfo{pages}{131} (\bibinfo{year}{2013}).

\bibitem[{\citenamefont{Hebeler et~al.}(2015)\citenamefont{Hebeler, Holt,
  Menendez, and Schwenk}}]{Hebe15ARNPS}
\bibinfo{author}{\bibfnamefont{K.}~\bibnamefont{Hebeler}},
  \bibinfo{author}{\bibfnamefont{J.~D.} \bibnamefont{Holt}},
  \bibinfo{author}{\bibfnamefont{J.}~\bibnamefont{Menendez}}, \bibnamefont{and}
  \bibinfo{author}{\bibfnamefont{A.}~\bibnamefont{Schwenk}},
  \bibinfo{journal}{Ann. Rev. Nucl. Part. Sci.} \textbf{\bibinfo{volume}{65}},
  \bibinfo{pages}{457} (\bibinfo{year}{2015}).

\bibitem[{\citenamefont{Navrátil et~al.}(2016)\citenamefont{Navrátil,
  Quaglioni, Hupin, Romero-Redondo, and Calci}}]{Navr16PS}
\bibinfo{author}{\bibfnamefont{P.}~\bibnamefont{Navrátil}},
  \bibinfo{author}{\bibfnamefont{S.}~\bibnamefont{Quaglioni}},
  \bibinfo{author}{\bibfnamefont{G.}~\bibnamefont{Hupin}},
  \bibinfo{author}{\bibfnamefont{C.}~\bibnamefont{Romero-Redondo}},
  \bibnamefont{and} \bibinfo{author}{\bibfnamefont{A.}~\bibnamefont{Calci}},
  \bibinfo{journal}{Phys. Scripta} \textbf{\bibinfo{volume}{91}},
  \bibinfo{pages}{053002} (\bibinfo{year}{2016}).

\bibitem[{\citenamefont{Hergert et~al.}(2013)\citenamefont{Hergert, Bogner,
  Binder, Calci, Langhammer et~al.}}]{Herg13IMSRG}
\bibinfo{author}{\bibfnamefont{H.}~\bibnamefont{Hergert}},
  \bibinfo{author}{\bibfnamefont{S.~K.} \bibnamefont{Bogner}},
  \bibinfo{author}{\bibfnamefont{S.}~\bibnamefont{Binder}},
  \bibinfo{author}{\bibfnamefont{A.}~\bibnamefont{Calci}},
  \bibinfo{author}{\bibfnamefont{J.}~\bibnamefont{Langhammer}},
  \bibnamefont{et~al.}, \bibinfo{journal}{Phys. Rev. C}
  \textbf{\bibinfo{volume}{87}}, \bibinfo{pages}{034307}
  (\bibinfo{year}{2013}).

\bibitem[{\citenamefont{Binder et~al.}(2014)\citenamefont{Binder, Langhammer,
  Calci, and Roth}}]{Bind14CCheavy}
\bibinfo{author}{\bibfnamefont{S.}~\bibnamefont{Binder}},
  \bibinfo{author}{\bibfnamefont{J.}~\bibnamefont{Langhammer}},
  \bibinfo{author}{\bibfnamefont{A.}~\bibnamefont{Calci}}, \bibnamefont{and}
  \bibinfo{author}{\bibfnamefont{R.}~\bibnamefont{Roth}},
  \bibinfo{journal}{Phys. Lett. B} \textbf{\bibinfo{volume}{736}},
  \bibinfo{pages}{119} (\bibinfo{year}{2014}).

\bibitem[{\citenamefont{Hergert et~al.}(2014)\citenamefont{Hergert, Bogner,
  Morris, Binder, Calci, Langhammer, and Roth}}]{Herg14MR}
\bibinfo{author}{\bibfnamefont{H.}~\bibnamefont{Hergert}},
  \bibinfo{author}{\bibfnamefont{S.~K.} \bibnamefont{Bogner}},
  \bibinfo{author}{\bibfnamefont{T.~D.} \bibnamefont{Morris}},
  \bibinfo{author}{\bibfnamefont{S.}~\bibnamefont{Binder}},
  \bibinfo{author}{\bibfnamefont{A.}~\bibnamefont{Calci}},
  \bibinfo{author}{\bibfnamefont{J.}~\bibnamefont{Langhammer}},
  \bibnamefont{and} \bibinfo{author}{\bibfnamefont{R.}~\bibnamefont{Roth}},
  \bibinfo{journal}{Phys. Rev. C} \textbf{\bibinfo{volume}{90}},
  \bibinfo{pages}{041302(R)} (\bibinfo{year}{2014}).

\bibitem[{\citenamefont{Hu et~al.}()\citenamefont{Hu, Zhang, Epelbaum,
  Mei{\ss}ner, and Meng}}]{Hu16nm5}
\bibinfo{author}{\bibfnamefont{J.}~\bibnamefont{Hu}},
  \bibinfo{author}{\bibfnamefont{Y.}~\bibnamefont{Zhang}},
  \bibinfo{author}{\bibfnamefont{E.}~\bibnamefont{Epelbaum}},
  \bibinfo{author}{\bibfnamefont{U.-G.} \bibnamefont{Mei{\ss}ner}},
  \bibnamefont{and} \bibinfo{author}{\bibfnamefont{J.}~\bibnamefont{Meng}},
  \eprint{arXiv:1612.05433}.

\bibitem[{\citenamefont{Hagen et~al.}(2016{\natexlab{a}})}]{Hage16NatPhys}
\bibinfo{author}{\bibfnamefont{G.}~\bibnamefont{Hagen}} \bibnamefont{et~al.},
  \bibinfo{journal}{Nature Phys.} \textbf{\bibinfo{volume}{12}},
  \bibinfo{pages}{186} (\bibinfo{year}{2016}{\natexlab{a}}).

\bibitem[{\citenamefont{Simonis et~al.}(2016)\citenamefont{Simonis, Hebeler,
  Holt, Men\'{e}ndez, and Schwenk}}]{Simo16unc}
\bibinfo{author}{\bibfnamefont{J.}~\bibnamefont{Simonis}},
  \bibinfo{author}{\bibfnamefont{K.}~\bibnamefont{Hebeler}},
  \bibinfo{author}{\bibfnamefont{J.~D.} \bibnamefont{Holt}},
  \bibinfo{author}{\bibfnamefont{J.}~\bibnamefont{Men\'{e}ndez}},
  \bibnamefont{and} \bibinfo{author}{\bibfnamefont{A.}~\bibnamefont{Schwenk}},
  \bibinfo{journal}{Phys. Rev. C} \textbf{\bibinfo{volume}{93}},
  \bibinfo{pages}{011302(R)} (\bibinfo{year}{2016}).

\bibitem[{\citenamefont{Simonis et~al.}(2017)\citenamefont{Simonis, Stroberg,
  Hebeler, Holt, and Schwenk}}]{Simo17SatFinNuc}
\bibinfo{author}{\bibfnamefont{J.}~\bibnamefont{Simonis}},
  \bibinfo{author}{\bibfnamefont{S.~R.} \bibnamefont{Stroberg}},
  \bibinfo{author}{\bibfnamefont{K.}~\bibnamefont{Hebeler}},
  \bibinfo{author}{\bibfnamefont{J.~D.} \bibnamefont{Holt}}, \bibnamefont{and}
  \bibinfo{author}{\bibfnamefont{A.}~\bibnamefont{Schwenk}},
  \bibinfo{journal}{Phys. Rev. C} \textbf{\bibinfo{volume}{96}},
  \bibinfo{pages}{014303} (\bibinfo{year}{2017}).

\bibitem[{\citenamefont{Drischler et~al.}(2016)\citenamefont{Drischler,
  Hebeler, and Schwenk}}]{Dris16asym}
\bibinfo{author}{\bibfnamefont{C.}~\bibnamefont{Drischler}},
  \bibinfo{author}{\bibfnamefont{K.}~\bibnamefont{Hebeler}}, \bibnamefont{and}
  \bibinfo{author}{\bibfnamefont{A.}~\bibnamefont{Schwenk}},
  \bibinfo{journal}{Phys. Rev. C} \textbf{\bibinfo{volume}{93}},
  \bibinfo{pages}{054314} (\bibinfo{year}{2016}).

\bibitem[{\citenamefont{Garcia~Ruiz et~al.}(2016)}]{Ruiz16Calcium}
\bibinfo{author}{\bibfnamefont{R.~F.} \bibnamefont{Garcia~Ruiz}}
  \bibnamefont{et~al.}, \bibinfo{journal}{Nature Phys.}
  \textbf{\bibinfo{volume}{12}}, \bibinfo{pages}{594} (\bibinfo{year}{2016}).

\bibitem[{\citenamefont{Hagen et~al.}(2016{\natexlab{b}})\citenamefont{Hagen,
  Jansen, and Papenbrock}}]{Hage16Ni78}
\bibinfo{author}{\bibfnamefont{G.}~\bibnamefont{Hagen}},
  \bibinfo{author}{\bibfnamefont{G.~R.} \bibnamefont{Jansen}},
  \bibnamefont{and}
  \bibinfo{author}{\bibfnamefont{T.}~\bibnamefont{Papenbrock}},
  \bibinfo{journal}{Phys. Rev. Lett.} \textbf{\bibinfo{volume}{117}},
  \bibinfo{pages}{172501} (\bibinfo{year}{2016}{\natexlab{b}}).

\bibitem[{\citenamefont{Birkhan et~al.}(2017)}]{Birk16edp}
\bibinfo{author}{\bibfnamefont{J.}~\bibnamefont{Birkhan}} \bibnamefont{et~al.},
  \bibinfo{journal}{Phys. Rev. Lett.} \textbf{\bibinfo{volume}{118}},
  \bibinfo{pages}{252501} (\bibinfo{year}{2017}).

\bibitem[{\citenamefont{Entem and Machleidt}(2003)}]{Ente03EMN3LO}
\bibinfo{author}{\bibfnamefont{D.~R.} \bibnamefont{Entem}} \bibnamefont{and}
  \bibinfo{author}{\bibfnamefont{R.}~\bibnamefont{Machleidt}},
  \bibinfo{journal}{Phys. Rev. C} \textbf{\bibinfo{volume}{68}},
  \bibinfo{pages}{041001(R)} (\bibinfo{year}{2003}).

\bibitem[{\citenamefont{Navr\'atil}(2007)}]{Navr07local3N}
\bibinfo{author}{\bibfnamefont{P.}~\bibnamefont{Navr\'atil}},
  \bibinfo{journal}{Few Body Syst.} \textbf{\bibinfo{volume}{41}},
  \bibinfo{pages}{117} (\bibinfo{year}{2007}).

\bibitem[{\citenamefont{Tews et~al.}(2016)\citenamefont{Tews, Gandolfi,
  Gezerlis, and Schwenk}}]{Tews16QMCPNM}
\bibinfo{author}{\bibfnamefont{I.}~\bibnamefont{Tews}},
  \bibinfo{author}{\bibfnamefont{S.}~\bibnamefont{Gandolfi}},
  \bibinfo{author}{\bibfnamefont{A.}~\bibnamefont{Gezerlis}}, \bibnamefont{and}
  \bibinfo{author}{\bibfnamefont{A.}~\bibnamefont{Schwenk}},
  \bibinfo{journal}{Phys. Rev. C} \textbf{\bibinfo{volume}{93}},
  \bibinfo{pages}{024305} (\bibinfo{year}{2016}).

\bibitem[{\citenamefont{Dyhdalo et~al.}(2016)\citenamefont{Dyhdalo, Furnstahl,
  Hebeler, and Tews}}]{Dyhd16Regs}
\bibinfo{author}{\bibfnamefont{A.}~\bibnamefont{Dyhdalo}},
  \bibinfo{author}{\bibfnamefont{R.~J.} \bibnamefont{Furnstahl}},
  \bibinfo{author}{\bibfnamefont{K.}~\bibnamefont{Hebeler}}, \bibnamefont{and}
  \bibinfo{author}{\bibfnamefont{I.}~\bibnamefont{Tews}},
  \bibinfo{journal}{Phys. Rev. C} \textbf{\bibinfo{volume}{94}},
  \bibinfo{pages}{034001} (\bibinfo{year}{2016}).

\bibitem[{\citenamefont{Coraggio et~al.}(2014)\citenamefont{Coraggio, Holt,
  Itaco, Machleidt, Marcucci, and Sammarruca}}]{Cora14nmat}
\bibinfo{author}{\bibfnamefont{L.}~\bibnamefont{Coraggio}},
  \bibinfo{author}{\bibfnamefont{J.~W.} \bibnamefont{Holt}},
  \bibinfo{author}{\bibfnamefont{N.}~\bibnamefont{Itaco}},
  \bibinfo{author}{\bibfnamefont{R.}~\bibnamefont{Machleidt}},
  \bibinfo{author}{\bibfnamefont{L.~E.} \bibnamefont{Marcucci}},
  \bibnamefont{and}
  \bibinfo{author}{\bibfnamefont{F.}~\bibnamefont{Sammarruca}},
  \bibinfo{journal}{Phys. Rev. C} \textbf{\bibinfo{volume}{89}},
  \bibinfo{pages}{044321} (\bibinfo{year}{2014}).

\bibitem[{\citenamefont{Sammarruca et~al.}(2015)\citenamefont{Sammarruca,
  Coraggio, Holt, Itaco, Machleidt, and Marcucci}}]{Samm15numat}
\bibinfo{author}{\bibfnamefont{F.}~\bibnamefont{Sammarruca}},
  \bibinfo{author}{\bibfnamefont{L.}~\bibnamefont{Coraggio}},
  \bibinfo{author}{\bibfnamefont{J.~W.} \bibnamefont{Holt}},
  \bibinfo{author}{\bibfnamefont{N.}~\bibnamefont{Itaco}},
  \bibinfo{author}{\bibfnamefont{R.}~\bibnamefont{Machleidt}},
  \bibnamefont{and} \bibinfo{author}{\bibfnamefont{L.~E.}
  \bibnamefont{Marcucci}}, \bibinfo{journal}{Phys. Rev. C}
  \textbf{\bibinfo{volume}{91}}, \bibinfo{pages}{054311}
  (\bibinfo{year}{2015}).

\bibitem[{\citenamefont{Schiavilla et~al.}(1998)}]{Schi98weakcap}
\bibinfo{author}{\bibfnamefont{R.}~\bibnamefont{Schiavilla}}
  \bibnamefont{et~al.}, \bibinfo{journal}{Phys. Rev. C}
  \textbf{\bibinfo{volume}{58}}, \bibinfo{pages}{1263} (\bibinfo{year}{1998}).

\bibitem[{\citenamefont{Raman et~al.}(1978)\citenamefont{Raman, Houser, and
  Walkiewicz}}]{Rama78tables}
\bibinfo{author}{\bibfnamefont{S.}~\bibnamefont{Raman}},
  \bibinfo{author}{\bibfnamefont{C.~A.} \bibnamefont{Houser}},
  \bibnamefont{and} \bibinfo{author}{\bibfnamefont{T.~A.}
  \bibnamefont{Walkiewicz}}, \bibinfo{journal}{At. Data Nucl. Data Tables}
  \textbf{\bibinfo{volume}{21}}, \bibinfo{pages}{567} (\bibinfo{year}{1978}).

\bibitem[{\citenamefont{Park et~al.}(2003)\citenamefont{Park, Marcucci,
  Schiavilla, Viviani, Kievsky, Rosati, Kubodera, Min, and
  Rho}}]{Park02eftsolar}
\bibinfo{author}{\bibfnamefont{T.~S.} \bibnamefont{Park}},
  \bibinfo{author}{\bibfnamefont{L.~E.} \bibnamefont{Marcucci}},
  \bibinfo{author}{\bibfnamefont{R.}~\bibnamefont{Schiavilla}},
  \bibinfo{author}{\bibfnamefont{M.}~\bibnamefont{Viviani}},
  \bibinfo{author}{\bibfnamefont{A.}~\bibnamefont{Kievsky}},
  \bibinfo{author}{\bibfnamefont{S.}~\bibnamefont{Rosati}},
  \bibinfo{author}{\bibfnamefont{K.}~\bibnamefont{Kubodera}},
  \bibinfo{author}{\bibfnamefont{D.~P.} \bibnamefont{Min}}, \bibnamefont{and}
  \bibinfo{author}{\bibfnamefont{M.}~\bibnamefont{Rho}},
  \bibinfo{journal}{Phys. Rev. C} \textbf{\bibinfo{volume}{67}},
  \bibinfo{pages}{055206} (\bibinfo{year}{2003}).

\bibitem[{\citenamefont{Ando et~al.}(2002)\citenamefont{Ando, Park, Kubodera,
  and Myhrer}}]{Ando02mud}
\bibinfo{author}{\bibfnamefont{S.}~\bibnamefont{Ando}},
  \bibinfo{author}{\bibfnamefont{T.~S.} \bibnamefont{Park}},
  \bibinfo{author}{\bibfnamefont{K.}~\bibnamefont{Kubodera}}, \bibnamefont{and}
  \bibinfo{author}{\bibfnamefont{F.}~\bibnamefont{Myhrer}},
  \bibinfo{journal}{Phys. Lett. B} \textbf{\bibinfo{volume}{533}},
  \bibinfo{pages}{25} (\bibinfo{year}{2002}).

\bibitem[{\citenamefont{Hoferichter et~al.}(2015)\citenamefont{Hoferichter,
  Klos, and Schwenk}}]{Hofe15powerdm}
\bibinfo{author}{\bibfnamefont{M.}~\bibnamefont{Hoferichter}},
  \bibinfo{author}{\bibfnamefont{P.}~\bibnamefont{Klos}}, \bibnamefont{and}
  \bibinfo{author}{\bibfnamefont{A.}~\bibnamefont{Schwenk}},
  \bibinfo{journal}{Phys. Lett. B} \textbf{\bibinfo{volume}{746}},
  \bibinfo{pages}{410} (\bibinfo{year}{2015}).

\bibitem[{\citenamefont{Entem et~al.}(2017)\citenamefont{Entem, Machleidt, and
  Nosyk}}]{Ente17EMn4lo}
\bibinfo{author}{\bibfnamefont{D.~R.} \bibnamefont{Entem}},
  \bibinfo{author}{\bibfnamefont{R.}~\bibnamefont{Machleidt}},
  \bibnamefont{and} \bibinfo{author}{\bibfnamefont{Y.}~\bibnamefont{Nosyk}},
  \bibinfo{journal}{Phys. Rev. C} \textbf{\bibinfo{volume}{96}},
  \bibinfo{pages}{024004} (\bibinfo{year}{2017}).

\bibitem[{\citenamefont{Nogga et~al.}(2003)\citenamefont{Nogga, Kievsky,
  Kamada, Gl{\"o}ckle, Marcucci, Rosati, and Viviani}}]{Nogg03triton}
\bibinfo{author}{\bibfnamefont{A.}~\bibnamefont{Nogga}},
  \bibinfo{author}{\bibfnamefont{A.}~\bibnamefont{Kievsky}},
  \bibinfo{author}{\bibfnamefont{H.}~\bibnamefont{Kamada}},
  \bibinfo{author}{\bibfnamefont{W.}~\bibnamefont{Gl{\"o}ckle}},
  \bibinfo{author}{\bibfnamefont{L.~E.} \bibnamefont{Marcucci}},
  \bibinfo{author}{\bibfnamefont{S.}~\bibnamefont{Rosati}}, \bibnamefont{and}
  \bibinfo{author}{\bibfnamefont{M.}~\bibnamefont{Viviani}},
  \bibinfo{journal}{Phys. Rev. C} \textbf{\bibinfo{volume}{67}},
  \bibinfo{pages}{034004} (\bibinfo{year}{2003}).

\bibitem[{\citenamefont{Epelbaum et~al.}(2005)\citenamefont{Epelbaum,
  Gl{\"o}ckle, and Mei{\ss}ner}}]{Epel05EGMN3LO}
\bibinfo{author}{\bibfnamefont{E.}~\bibnamefont{Epelbaum}},
  \bibinfo{author}{\bibfnamefont{W.}~\bibnamefont{Gl{\"o}ckle}},
  \bibnamefont{and} \bibinfo{author}{\bibfnamefont{U.-G.}
  \bibnamefont{Mei{\ss}ner}}, \bibinfo{journal}{Nucl. Phys. A}
  \textbf{\bibinfo{volume}{747}}, \bibinfo{pages}{362} (\bibinfo{year}{2005}).

\bibitem[{\citenamefont{Hoppe et~al.}()\citenamefont{Hoppe, Drischler,
  Furnstahl, Hebeler, and Schwenk}}]{Hopp17WeinEVAn}
\bibinfo{author}{\bibfnamefont{J.}~\bibnamefont{Hoppe}},
  \bibinfo{author}{\bibfnamefont{C.}~\bibnamefont{Drischler}},
  \bibinfo{author}{\bibfnamefont{R.~J.} \bibnamefont{Furnstahl}},
  \bibinfo{author}{\bibfnamefont{K.}~\bibnamefont{Hebeler}}, \bibnamefont{and}
  \bibinfo{author}{\bibfnamefont{A.}~\bibnamefont{Schwenk}},
  \eprint{arXiv:1707.06438}.

\bibitem[{\citenamefont{Epelbaum}(2006{\natexlab{b}})}]{Epel06PPNP}
\bibinfo{author}{\bibfnamefont{E.}~\bibnamefont{Epelbaum}},
  \bibinfo{journal}{Prog. Part. Nucl. Phys.} \textbf{\bibinfo{volume}{57}},
  \bibinfo{pages}{654} (\bibinfo{year}{2006}{\natexlab{b}}).

\bibitem[{\citenamefont{Carbone et~al.}(2014)\citenamefont{Carbone, Rios, and
  Polls}}]{Carb14SCGFdd}
\bibinfo{author}{\bibfnamefont{A.}~\bibnamefont{Carbone}},
  \bibinfo{author}{\bibfnamefont{A.}~\bibnamefont{Rios}}, \bibnamefont{and}
  \bibinfo{author}{\bibfnamefont{A.}~\bibnamefont{Polls}},
  \bibinfo{journal}{Phys. Rev. C} \textbf{\bibinfo{volume}{90}},
  \bibinfo{pages}{054322} (\bibinfo{year}{2014}).

\bibitem[{\citenamefont{Dutra et~al.}(2012)\citenamefont{Dutra,
  Louren\ifmmode~\mbox{\c{c}}\else \c{c}\fi{}o, S\'a~Martins, Delfino, Stone,
  and Stevenson}}]{Dutra12skyrme}
\bibinfo{author}{\bibfnamefont{M.}~\bibnamefont{Dutra}},
  \bibinfo{author}{\bibfnamefont{O.}~\bibnamefont{Louren\ifmmode~\mbox{\c{c}}\else
  \c{c}\fi{}o}}, \bibinfo{author}{\bibfnamefont{J.~S.}
  \bibnamefont{S\'a~Martins}},
  \bibinfo{author}{\bibfnamefont{A.}~\bibnamefont{Delfino}},
  \bibinfo{author}{\bibfnamefont{J.~R.} \bibnamefont{Stone}}, \bibnamefont{and}
  \bibinfo{author}{\bibfnamefont{P.~D.} \bibnamefont{Stevenson}},
  \bibinfo{journal}{Phys. Rev. C} \textbf{\bibinfo{volume}{85}},
  \bibinfo{pages}{035201} (\bibinfo{year}{2012}).

\bibitem[{\citenamefont{Epelbaum
  et~al.}(2015{\natexlab{b}})\citenamefont{Epelbaum, Krebs, and
  Mei{\ss}ner}}]{Epel15improved}
\bibinfo{author}{\bibfnamefont{E.}~\bibnamefont{Epelbaum}},
  \bibinfo{author}{\bibfnamefont{H.}~\bibnamefont{Krebs}}, \bibnamefont{and}
  \bibinfo{author}{\bibfnamefont{U.-G.} \bibnamefont{Mei{\ss}ner}},
  \bibinfo{journal}{Eur. Phys. J. A} \textbf{\bibinfo{volume}{51}},
  \bibinfo{pages}{53} (\bibinfo{year}{2015}{\natexlab{b}}).

\bibitem[{\citenamefont{Binder et~al.}(2016)}]{Bind15Fewbody}
\bibinfo{author}{\bibfnamefont{S.}~\bibnamefont{Binder}} \bibnamefont{et~al.},
  \bibinfo{journal}{Phys. Rev. C} \textbf{\bibinfo{volume}{93}},
  \bibinfo{pages}{044002} (\bibinfo{year}{2016}).

\bibitem[{\citenamefont{Carbone et~al.}(2013)\citenamefont{Carbone, Cipollone,
  Barbieri, Rios, and Polls}}]{Carb13SCGF3B}
\bibinfo{author}{\bibfnamefont{A.}~\bibnamefont{Carbone}},
  \bibinfo{author}{\bibfnamefont{A.}~\bibnamefont{Cipollone}},
  \bibinfo{author}{\bibfnamefont{C.}~\bibnamefont{Barbieri}},
  \bibinfo{author}{\bibfnamefont{A.}~\bibnamefont{Rios}}, \bibnamefont{and}
  \bibinfo{author}{\bibfnamefont{A.}~\bibnamefont{Polls}},
  \bibinfo{journal}{Phys. Rev. C} \textbf{\bibinfo{volume}{88}},
  \bibinfo{pages}{054326} (\bibinfo{year}{2013}).

\bibitem[{\citenamefont{Valderrama and Phillips}(2015)}]{Vald15pcCurr}
\bibinfo{author}{\bibfnamefont{M.~P.} \bibnamefont{Valderrama}}
  \bibnamefont{and} \bibinfo{author}{\bibfnamefont{D.~R.}
  \bibnamefont{Phillips}}, \bibinfo{journal}{Phys. Rev. Lett.}
  \textbf{\bibinfo{volume}{114}}, \bibinfo{pages}{082502}
  (\bibinfo{year}{2015}).

\end{thebibliography}

\newpage

\setcounter{footnote}{0}

\onecolumngrid
\begin{center}
{\large\bf Erratum: Uncertainties in constraining low-energy constants from $^3$H $\beta$ decay\\
{[}Eur. Phys. J. A 53, 168 (2017){]}}
\bigskip

P.\ Klos,\textsuperscript{1,2,}\footnote{pklos@theorie.ikp.physik.tu-darmstadt.de} A.\ Carbone,\textsuperscript{1,2,}\footnote{arianna@theorie.ikp.physik.tu-darmstadt.de} K.\ Hebeler,\textsuperscript{1,2,}\footnote{kai.hebeler@physik.tu-darmstadt.de} J.\ Men{\'e}ndez,\textsuperscript{3,}\footnote{menendez@nt.phys.s.u-tokyo.ac.jp} and A.\ Schwenk\textsuperscript{1,2,4,}\footnote{schwenk@physik.tu-darmstadt.de}
\smallskip

{\small\it
\textsuperscript{1}Institut f\"ur Kernphysik, 
Technische Universit\"at Darmstadt, 
64289 Darmstadt, Germany 

\textsuperscript{2}ExtreMe Matter Institute EMMI, 
GSI Helmholtzzentrum f\"ur Schwerionenforschung GmbH, 
64291 Darmstadt, Germany

\textsuperscript{3}Department of Physics, University of Tokyo, Hongo, 
Tokyo 113-0033, Japan

\textsuperscript{4}Max-Planck-Institut f\"ur Kernphysik, 
Saupfercheckweg 1, 
69117 Heidelberg, 
Germany}
\end{center}
\bigskip

\twocolumngrid

\setcounter{figure}{1} 
In our article Eur.\ Phys.\ J.\ {\bf A}\ 53, 168\ (2017) the low-energy constant $c_D$ in the two-body currents was used incorrectly. The correct definition, which is consistent with the $c_D$ coupling in the leading three-nucleon forces, gives for Eq.~\eqref{eq:dR_def}:

\begin{equation} d_R=-\frac{1}{4\Lambda_\chi
g_A}c_D+\frac{1}{3}(c_3+2c_4)+\frac{1}{6m_N}\,.
\end{equation}
As a result the $c_D$ dependence of the Gamow-Teller matrix elements has changed. Here we provide the corrected Figs.~2--6. While the curves change significantly due to the $-\frac{1}{4}$ change in the $c_D$ part of the two-body currents, we emphasize that the discussion provided in our original manuscript remains qualitatively correct, as there is still a strong dependence of the results on the value of $c_D$.

We thank R.\ Schiavilla for pointing out the inconsistent definition of $c_D$, and E.\ Epelbaum and H.\ Krebs for discussions.

\vspace{5cm}

\begin{figure}[htb]
\begin{center}
\includegraphics[width=0.48\textwidth,clip=]{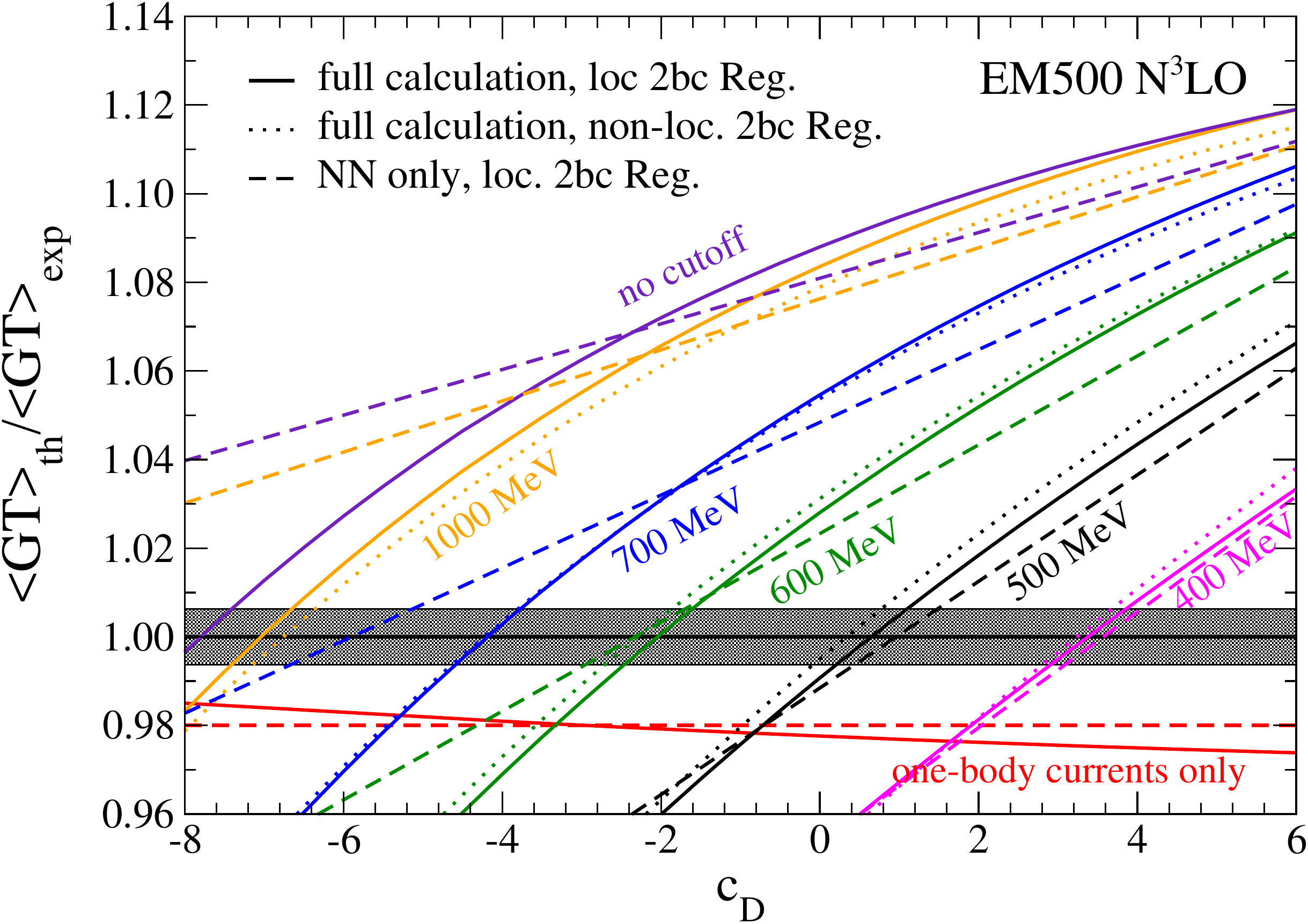}
\end{center}
\caption{(Color online) Ratio of calculated and experimental Gamow-Teller
matrix elements as a function of $c_D$ for different cutoff values
and regulators in the two-body currents, based on
the EM 500 MeV N$^3$LO potential of Ref.~[44].
The solid (dotted) lines show results for nuclear states including 3N forces at
N$^2$LO using the $^3$H binding energy constraint of Fig.~1 for
a local [non-local with $n = 2$, see Eq.(8)] regulator in the two-body currents. For comparison, we
also show results based on NN interactions only (dashed lines) and with 1b
currents only.  The width of the shaded band denotes the $2\sigma$
experimental uncertainty. }
\end{figure}

\newpage
\onecolumngrid

\begin{figure}[htb]
\begin{minipage}[t]{0.48\columnwidth}
\begin{center}
\includegraphics[width=\textwidth,clip=]{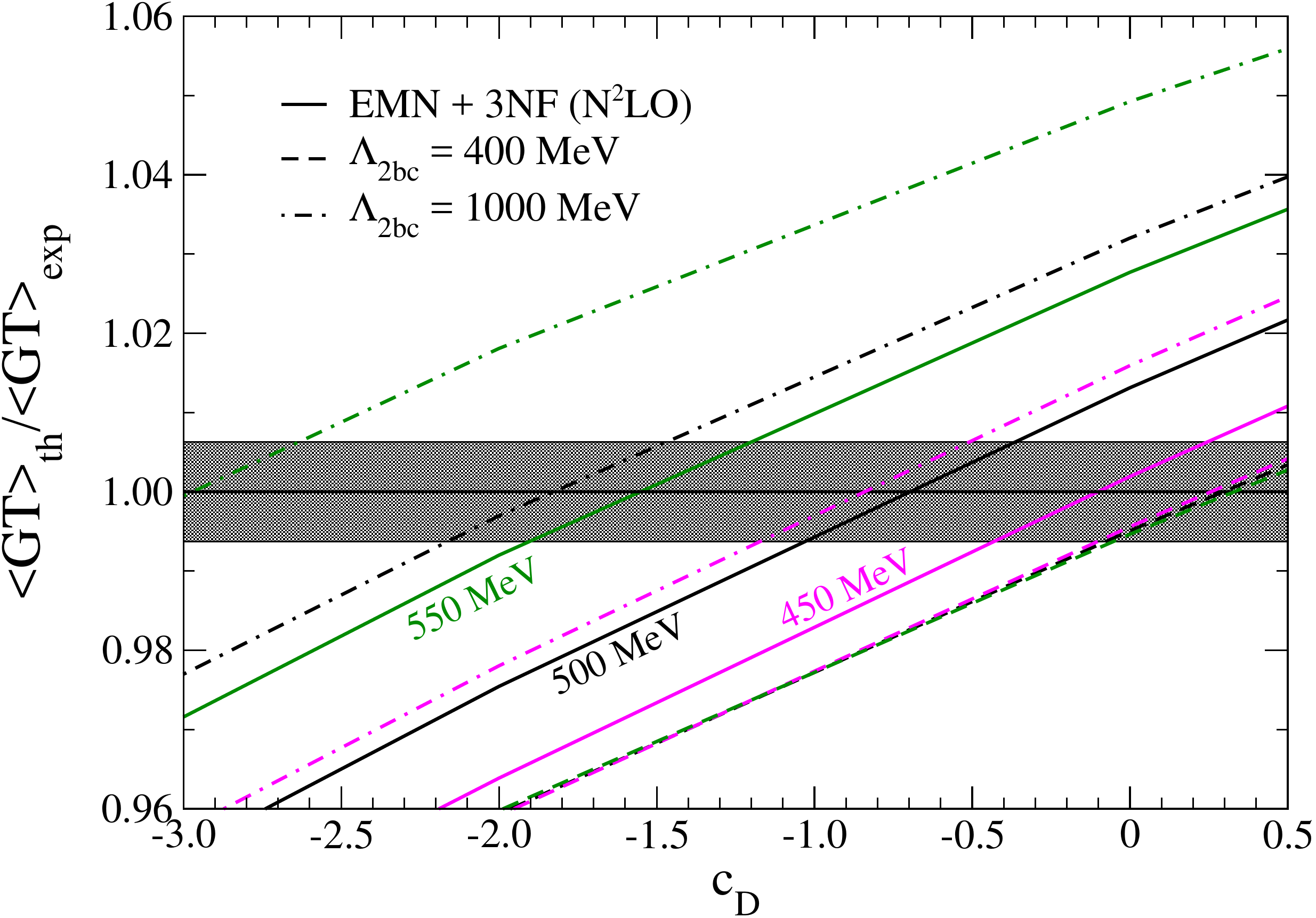}
\end{center}
\caption{(Color online) Ratio of calculated and experimental
Gamow-Teller matrix elements as a function of $c_D$ based on the EMN
potentials at order N$^2$LO of Ref.~[55]. We show results for the 2b current cutoff $\Lambda_\text{2bc}=400$~MeV (dashed lines) and $\Lambda_\text{2bc}=1000$~MeV (dash-dotted lines)
using a non-local regulator [see Eq.(8)] with
$n = 2$, whereas the solid lines show the cases 
when using the same cutoff values in the regulators for the interactions and currents.
The width of the shaded band denotes the $2\sigma$ experimental uncertainty.}
\vspace{0.5cm}
\begin{center}
\includegraphics[width=\textwidth,clip=]{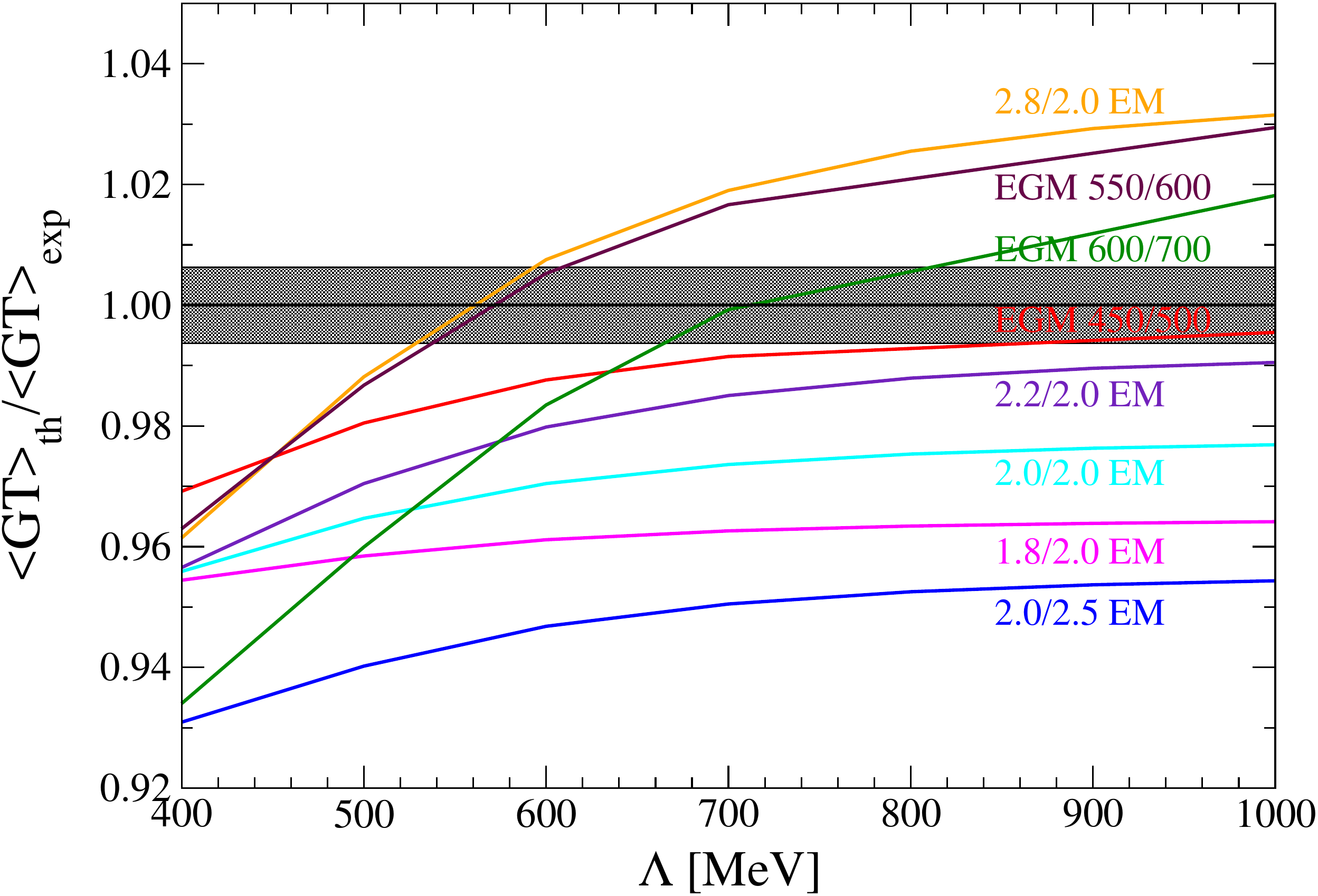}
\end{center}
\caption{(Color online) Ratio of calculated and experimental Gamow-Teller
matrix elements as a function of the cutoff $\Lambda$ in the non-local
regulator [$n=2$, see Eq.~(8)] for a set of chiral
interactions, using different fitting observables: the results labeled 'EM'
are based on NN plus 3N interactions, for which the $c_D$ and $c_E$ values are
fit to the binding energy of $^3$H and the charge radius of $^4$He (see
Ref.~[26] for details). The results labeled 'EGM' are based on NN
plus 3N forces fitted to the binding energy of $^3$H and the neutron-deuteron
scattering length~[59]. The width of the shaded band denotes the
$2\sigma$ experimental uncertainty.}
\end{minipage}
\hfill
\begin{minipage}[t]{0.48\columnwidth}
\begin{center}
\includegraphics[width=\textwidth,clip=]{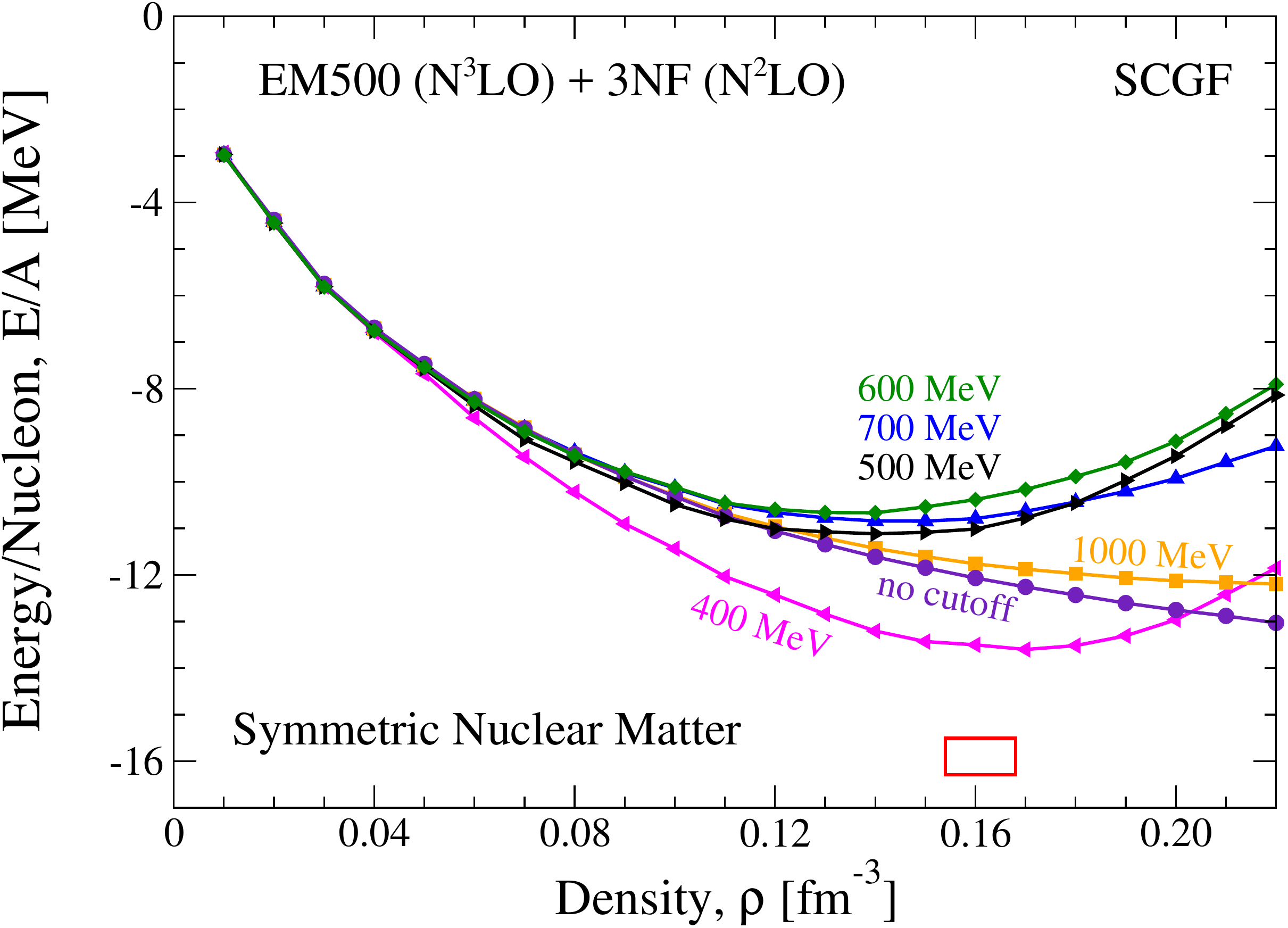}
\end{center}
\caption{(Color online) Energy per nucleon of symmetric nuclear matter as a
function of nucleon density obtained within the self-consistent Green's
function approach~[60]. Results are based on the NN EM 500 MeV
at N$^3$LO, including normal-ordered 3N interaction contributions at N$^2$LO.
The curves correspond to different $c_D$ and $c_E$ values obtained according
to Figs.~1 and~\ref{Fig_cD_Reg_dep}. The box describes the
range for the empirical saturation point provided by mean-field
calculations~[61].}
\vspace{0.5cm}
\begin{center}
\includegraphics[width=\textwidth,clip=]{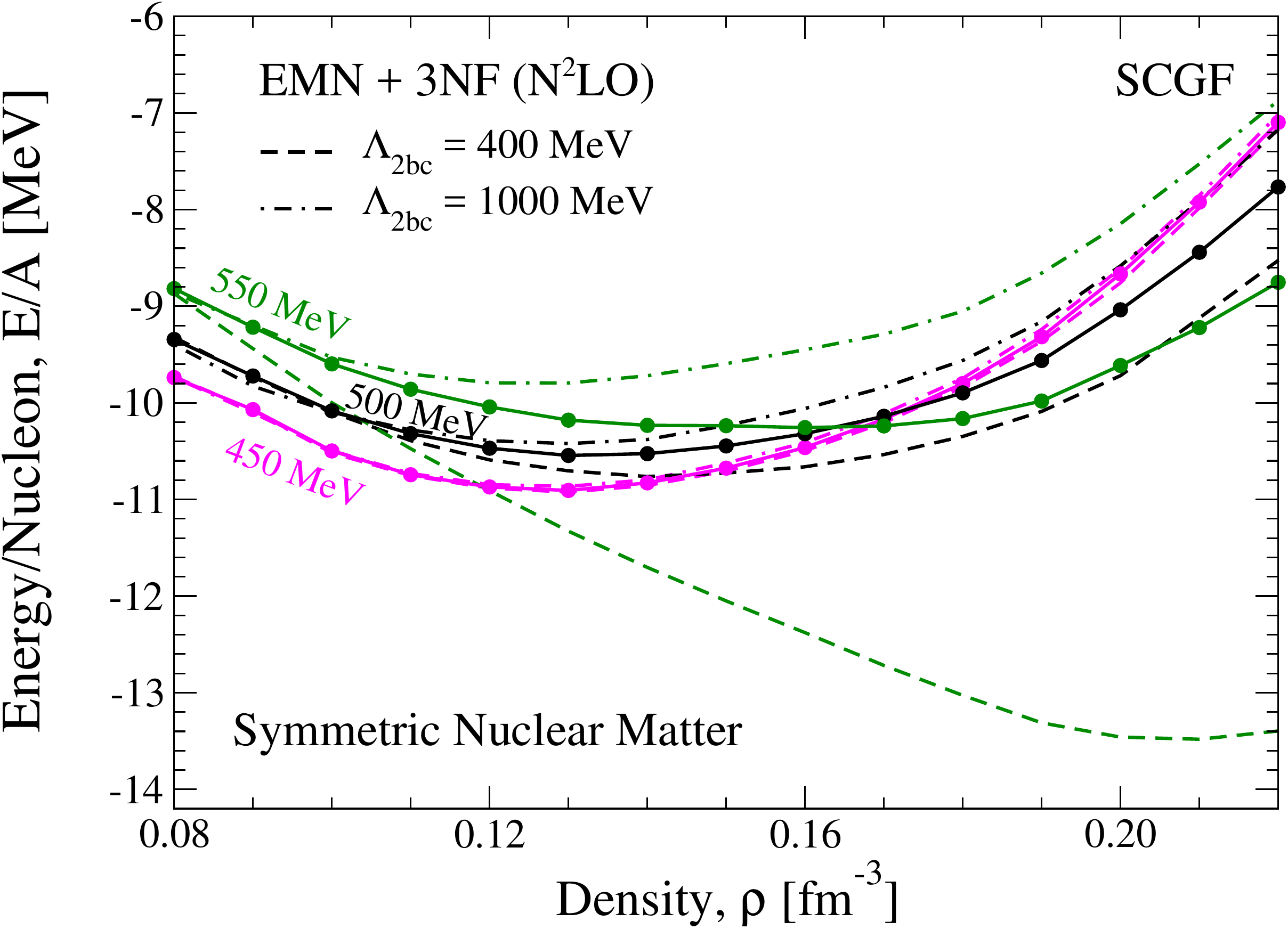}
\end{center}
\caption{(Color online) Same as Fig.~\ref{Fig_snm_ener} but
showing results based on the EMN potentials at N$^2$LO. The curves correspond to
different $c_D$ and $c_E$ values obtained according to Figs.~1
and~\ref{Fig_cD_Reg_dep_EMN}.
The different lines correspond to 2b currents cutoffs equivalent to the interaction (solid), $\Lambda_\text{2bc}=400$~MeV (dashed), and $\Lambda_\text{2bc}=1000$~MeV (dash-dotted) as  shown in Fig.~\ref{Fig_cD_Reg_dep_EMN}.}
\end{minipage}
\end{figure}

\end{document}